\documentclass{aastex631}

\usepackage{graphicx}
\usepackage{amsmath}
\usepackage{xcolor}
\usepackage{color, soul}
\usepackage{placeins}
\hypersetup{colorlinks=true, allcolors=blue}

\begin{document}

\title{Evolution of Coronal Mass Ejection Properties through Superposed Epoch Analysis from 0.2 to 2.2 au}

\author{Yakub Olufadi}
\affiliation{Space Science Center and Department of Physics, University of New Hampshire, Durham, NH 03284, USA.}

\author{Nada Al-Haddad}
\affiliation{Space Science Center and Department of Physics, University of New Hampshire, Durham, NH 03284, USA.}

\author{Florian Regnault}
\affiliation{Space Science Center and Department of Physics, University of New Hampshire, Durham, NH 03284, USA.}

\author{No{\'e} Lugaz}
\affiliation{Space Science Center and Department of Physics, University of New Hampshire, Durham, NH 03284, USA.}

\author{Bin Zhuang}
\affiliation{Space Science Center and Department of Physics, University of New Hampshire, Durham, NH 03284, USA.}

\author{Charles~J.  Farrugia}
\affiliation{Space Science Center and Department of Physics, University of New Hampshire, Durham, NH 03284, USA.}

\author{Christian M{\"o}stl}
\affiliation{Austrian Space Weather Office, GeoSphere Austria, Graz, Austria.}

\author{Emma E. Davies}
\affiliation{Austrian Space Weather Office, GeoSphere Austria, Graz, Austria.}

\author{Eva Weiler}
\affiliation{Austrian Space Weather Office, GeoSphere Austria, Graz, Austria.}
\affiliation{Institute of Physics, University of Graz, Graz, Austria.}



\begin{abstract}
Coronal mass ejections (CMEs) are explosive and energetic events consisting of strong magnetic structures erupting from the solar corona. We use superposed epoch analysis to investigate the general properties of CMEs as measured {\it in situ} from 0.2 to 2.2 au based on over 1600 events obtained from the HELIO4CAST catalog. We examine the dependence of the CME global properties on solar cycle phase, and compare the CME parameters derived in the active phase (AP) with the quiet phase (QP).
Our findings show that during the AP of the solar cycle, the occurring CMEs are faster and have stronger magnetic field strength than during the QP, which has denser but  weaker magnetic strength. These differences in magnetic field strength and density remain even when controlling for the speed.
This may indicate that the enhanced profiles observed during the AP are not only a consequence of the CME propagation speed but may also reflect intrinsic differences in the eruption mechanism during different solar cycle phases.
We also study how the magnetic field strength and components of the CME magnetic ejecta (ME) structure evolve with heliocentric distance. We find that the toroidal and poloidal ME magnetic field components have a similar power law decrease with distance, indicating a comparable expansion behavior of CMEs in these dimensions. 
We further quantify the CME magnetic field asymmetry 
using the front-to-rear ratio of the toroidal component across heliocentric distance and find evidence of an increase of this ratio with heliocentric distance.

\end{abstract}

\keywords{Coronal Mass Ejections, Solar Cycle Dependency, Heliocentric Distance Dependency}

\section{Introduction} \label{sec:intro}

First directly observed in 1971 \citep{Tousey1973,Howard1982}, coronal mass ejections (CMEs) were initially observed as strong explosions in visible (white) light, which were later recognized as major drivers of natural hazards \citep{Alexander2006, Cremades2007, Gopalswamy2016} and associated with unusual interplanetary signatures \citep[e.g., see][]{Gosling1973}. CMEs and their associated shock waves give rise to solar energetic particle (SEP) events, as well as geomagnetic storms \citep{Wilson1987, Gosling1993}. Although CMEs can be studied remotely through coronagraph imaging or via {\it in~situ} measurements, in this work we use {\it in~situ} data measurements to study the CMEs properties as they pass spacecraft in various locations of the inner heliosphere. We use the term CME to describe these eruptions no matter whether they are measured remotely or {\it in situ}.

When CMEs are measured {\it in situ}, the entire CME structure can be divided into substructures: the sheath which is a region of heated and compressed solar wind plasma often bounded by a shock front along with a highly varying magnetic field \citep{Gosling1990, Kilpua2017, Salman2020}, and a magnetic ejecta (ME) region which is defined by a strong magnetic field strength with or without a smooth magnetic field rotation \citep{Burlaga1981, Wimmer-Schweingruber2006}. 

MEs typically have low proton temperature, low proton $\beta$ (the ratio of thermal to magnetic pressures) \citep{Gosling1973, Richardson1995, Wang2005}, lower proton density, increased ion charge state \citep{Wimmer-Schweingruber2006, Zurbuchen2006}, abundant helium, and counter-streaming suprathermal ($>$80 eV) electron beams \citep{Borrini1982, Gosling1986}; however, their exact nature and properties are still an active area of research \citep[see recent review by][]{Al-Haddad2025}. 

The properties of CMEs in the corona vary with solar activity \citep{Cremades2007, Vourlidas2010,Jones2020}. Several other studies have demonstrated these effects, for example, \citet{Hundhausen1999} found that CMEs erupting during quiet periods tend to be slower than those emitted during active periods. Likewise, \citet{Song2021} showed that the ionic charge states of CMEs are positively correlated with sunspot numbers, confirming a connection between solar activity and CME composition. \citet{Regnault2020} analyzed 20 years of {\it in~situ} CME observations from ACE to investigate how the internal structure of CMEs varies between active and quiet phases, and their findings confirmed that such variations do exist.

Studying CME properties such as magnetic field strength at various heliocentric distances is essential for understanding CME evolution \citep[]{Bothmer1998,Liu2005}. For example, \citet{Janvier2019} showed that the magnetic field strength profiles of CMEs within the sheath and ME exhibit correlations with their propagation velocities. In addition, \citet{Winslow2015}, \citet{Good2016}, and \citet{Davies2021} conducted statistical analyses of CME properties over different heliocentric distance ranges using measurements  from planetary missions and spacecraft near 1~au to investigate how the average and maximum magnetic field strengths in the ME vary with heliocentric distance. Their findings are consistent with earlier findings from \citet{Gulisano2010} using Helios measurements, as both studies observed a similar power law dependence of the maximum magnetic field strength in the ME. However, \citet{Lugaz2020} following up work by \citet{Salman2020a}, and, for a different dataset, \citet{Davies2022} found that the CME global expansion rate, estimated by the decrease in magnetic field strength with distance, is inconsistent with local expansion properties, measured by the bulk speed decrease near 1~au.

Meanwhile, the knowledge of how ME magnetic field components evolve during expansion is important because the total magnetic field strength alone does not capture how individual magnetic field components decay during propagation. It is also essential to understand if the ME evolves towards or away from a force-free state, for example. The variations in the axial and azimuthal components can reveal whether the CME expands isotropically or whether certain directions stretch more rapidly than others, leading to anisotropic behavior. \citet{Yu2024} investigated this by multi-spacecraft measurements for radially aligned conjunction events. Their analysis showed that the local expansion of the ME appeared anisotropic, whereas the overall global expansion of the ME structure is largely isotropic. However, their findings were based on limited event statistics, as such events with two or more spacecraft in conjunction are rare. As a result, the broader statistical behavior of axial (toroidal) and azimuthal (poloidal) component evolution remains unclear. In this study, we investigate a large sample of CMEs to statistically examine the expansion characteristics of the toroidal and poloidal field components within the ME across different heliocentric distances. We further assess the CME magnetic field asymmetry by comparing the front and rear toroidal field components as a consequence of CME evolution during propagation. This front-to-rear asymmetry can be thought of as a consequence of the CME aging \citep[]{Farrugia1992, Osherovich1993, Regnault2023b, Regnault2024}.

Furthermore, we examine the dependence of CME properties on the phase of the solar cycle, specifically comparing active and quiet phases, using nearly three decades (1995--2023) of {\it in~situ} measurements from multiple spacecraft (STEREO-A, STEREO-B, and Wind) near 1~au. Unlike \citet{Regnault2020}, who relied on a single spacecraft (ACE), our multi-spacecraft approach provides a substantially larger and more diverse dataset that spans multiple solar cycles and includes a richer set of plasma and magnetic field parameters. This enables a more statistically robust validation and extension of their findings from a broader, multi-mission perspective. To the best of our knowledge, this is one of the first studies to examine this effect using multi-mission {\it in~situ} measurements not only for fundamental CME parameters but also for derived quantities such as the Alfv{\'e}n Mach number and ram pressure, which provide physical insight into CME and solar wind interactions.

With the foci of solar cycle activity dependence and heliocentric distance dependence, we investigate over 1000 CMEs in the inner heliosphere recorded in the HELIO4CAST catalog \citep{Christian2017,Christian2020, Christian2026} by adopting the Superposed Epoch Analysis (SEA) method. The rest of this paper is organized as follows: Section~2 presents the dataset, the methods and the categorization of CMEs used in this study. Section~3 discusses the solar cycle dependence of CME properties: active phase (AP) and quiet phase (QP) CMEs. We discuss the CME heliocentric distance dependency in Section~4; and finally we discuss the results and conclude in Section~5.

\section{Data and Method}
\subsection{CME Catalog and Dataset}

For this study, we used nearly three decades of interplanetary probe measurements from multiple spacecraft, identified by the HELIO4CAST ICMECAT version~2.1 released on 01-September-2023 \citep{Christian2017,Mostl2020, Christian2026}. This catalog is an initiative by the Austrian Space Weather Office within the GeoSphere Austria, to maintain a living catalog of {\it in~situ} observations of CMEs. This catalog has been used to build a machine learning model to forecast the minimum $B_z$ component in CMEs \citep{Reiss2021}, examine the CME events causing the May 2024 superstorm and its effect on Earth's magnetosphere \citep{Weiler2025}, vet the magnetic field structure and propagation of CMEs with a three-dimensional flux rope model \citep{Weiss2021, weiss_2024}, and train artificial intelligence algorithms for automatically detecting CMEs with {\it in~situ} observations \citep{Rudisser2026}, among others.

The HELIO4CAST catalog comprises the event boundaries (timestamps) for both magnetic field data and plasma (proton) data measurements (when available) from approximately 0.1~au to over 5.0~au, from multiple spacecraft including Solar Orbiter (2020--present), Parker Solar Probe (PSP, 2018--Present), STEREO-A (2006--present), STEREO-B (2006--2014), Venus Express (2006--2014), MESSENGER (2004--2015), Juno cruise phase (2011--2016), and {\it Wind} (1994--present). The catalog provides CME start time, ME start and end times (MEs are referred to as magnetic obstacles, MOs, in the catalog). If there is a shock, then the CME start time is the shock time; if not, then the CME start time is the ME time. It also lists the heliocentric distances of the measurements and contains over a thousand events when this study was started. STEREO-A, STEREO-B, and Wind are near 1~au \citep{Stone1998, Kaiser2008}, Solar Orbiter covers heliocentric distances ranges between 0.28~au and 1.2~au \citep{Muller2020}, and PSP  between 0.046~au and 1~au \citep{Raouafi2023}. MESSENGER \citep{Solomon2001}, Venus Express \citep[VEX;][]{Zhang2006}, and Juno \citep{Connerney2017} took measurements between 0.31~au and 0.47~au; 0.72~au and 0.73~au; 1.0~au and 5~au respectively, and only have magnetic field measurements. We used the \cite{Winslow2013} catalog to locate Mercury's bow shock and removed magnetospheric crossings in MESSENGER data and also removed any CME events with data gaps for all missions involved in this work.

We focus on the events from January 1995 to March 2023. We resample all the data into 1-minute resolution. The coordinate system for all magnetic field components follows the heliocentric radial-tangential-normal (RTN) coordinate system, where the R~axis points directly away from the Sun to the spacecraft, the T~axis lies in the plane of the Sun's equator, perpendicular to the radial direction, pointing in the direction of planetary (or spacecraft) motion, and the N~axis is perpendicular to both R and T, pointing northward out of the plane of the Sun's equator, completing a right-handed coordinate system.

\subsection{Superposed Epoch Analysis (SEA)}

Superposed Epoch Analysis (SEA) is a statistical method used to detect patterns or trends in time-series data by aligning events based on a reference time (epoch) \citep{Chree1913}. A commonly used extension of this method involves normalizing the time axis between two or more boundary times, allowing events of different durations to be compared on a common temporal scale (e.g., \citealt{Mas-Meza2016}). The SEA was applied to a large number of CMEs involved in this study; for solar cycle CME dependency and heliocentric distance dependency, with time series of the different {\it in~situ} physical parameters. Doing so allows us to investigate the global physical properties of the CMEs and permits the observation of the temporal profile variation across different CME substructures.
Several studies have utilized the SEA method for their analysis; for instance, the geomagnetic storms were categorized into intense and onset types \citep{Hutchinson2011}, the investigation of the propagation of CMEs in the inner heliosphere \citep{Janvier2019}, the study of CME substructures near Earth and their effects on galactic cosmic rays \citep{Mas-Meza2016}, characterize CME-related Forbush decreases at Mercury \citet{Davies2023}, the investigation of average small flux-rope magnetic field profiles \citep{Banu2024}, and the identification of generic features in interplanetary CMEs \citep{Regnault2020}.


We normalized the CME events in time prior to performing the SEA to account for their inherently different durations and to ensure the same event duration for all the CMEs. This normalization was applied separately to each CME substructure rather than to the entire CME duration. Specifically, the sheath and ME intervals were independently rescaled in time. We adopted median durations of 7.75~hours for the sheath and 22.28~hours for the ME, obtained from the ensemble duration distributions of CME events observed by {\it Wind}, STEREO-A, and STEREO-B.

To better compute a complete profile of CME substructures, including its surrounding solar wind, we define two other time intervals or boundaries: the pre-CME solar wind (pre-SW) and the post-CME solar wind (the wake). The time span chosen for these substructures corresponds to half the sum of the sheath and ME durations ($\sim15\,\text{hr})$ . Then, we bin all CME parameters over time, resulting in a uniform number of time points for every event. 

Generally, the SEA results are summarized by either the mean of the distribution or its median in each time bin \citep{Mas-Meza2016, Rodriguez2016, Janvier2019, Yermolaev2020, Regnault2020}. So in this study, we present the SEA results using both the mean and median distribution profiles.

\subsection{Minimum Variance Analysis (MVA)} \label{subsec:MVA}

In this work we aim to study the variation of magnetic field components in axial (the component of the magnetic field along the central axis of the flux rope) and poloidal (the helical component) directions. To do so, we perform a minimum variance analysis (MVA) on all CMEs. The MVA method \citep{Sonnerup1967} involves the computation of eigenvectors and eigenvalues from a covariance matrix of magnetic field components and creates a coordinate system for each event, with one axis pointing in the direction of the least variance (``min''). The two other axes are the intermediate (``int'') or axial/toroidal component ($B_{k}$) axis and maximally (``max'') varying or poloidal ($B_{j}$) directions. Once the coordinate system has been identified, the magnetic field components within that system can be calculated. 

For better selection of the robust MVA events in each bin, we utilize the ratio of intermediate-to-minimum eigenvalues ($r_{\mathrm{ratio}}$) and set a threshold of $r_{\mathrm{ratio}} > 2$, retaining only events that meet this criterion consistent with previous studies \citep{Cartwright2010, Yu2024, Banu2024}. We then use the eigenvectors to transform the magnetic field components originally in the RTN coordinate systems to the MVA coordinates, from which we obtain field variation plots of the $B_{k}$ and $B_{j}$ components. 

Because MVA eigenvectors are defined only up to a sign, the derived $B_{j}$ and $B_{k}$ components may appear reversed in polarity from event to event, even when the underlying flux-rope structure is physically comparable. To ensure a consistent sign convention for the MVA components, we fit a straight line to the event time series of $B_{\phi}$ and multiply the entire profile by $-1$ whenever the fitted slope is negative; otherwise, the original sign is retained. Similarly, for $B_{z}$, the entire profile is multiplied by $-1$ whenever the event-averaged $B_{z}$ is negative; otherwise, the original sign is retained.

\begin{figure}
    \centering
    \includegraphics[width=\textwidth]{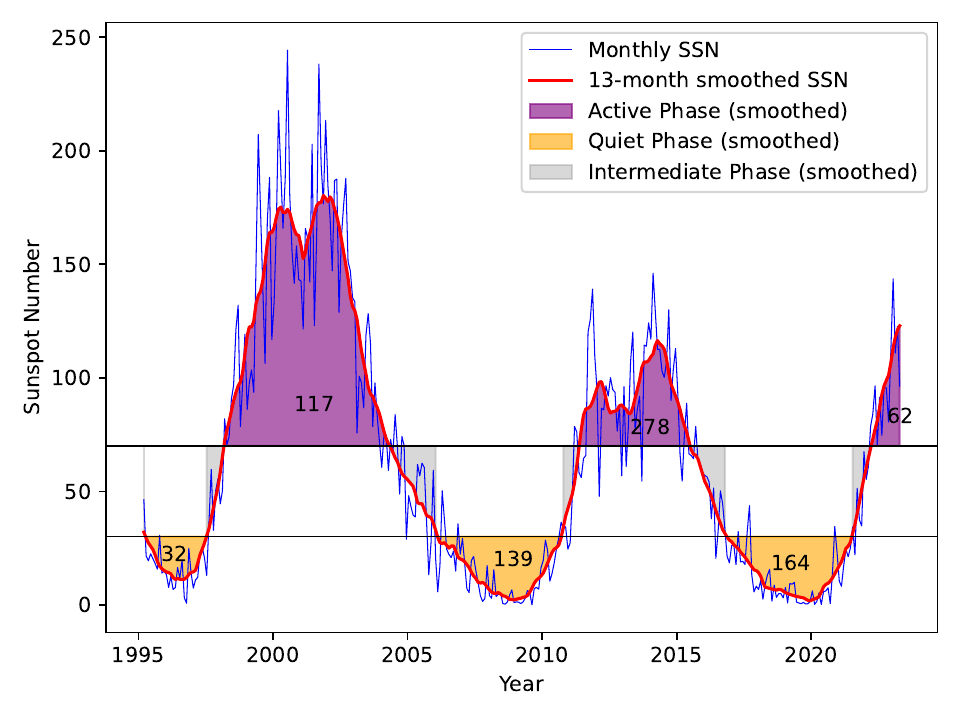}
    \caption{The (monthly and 13-month smoothed) SSN plot describes the active (purple) and quiet (orange) phases of solar cycle in which the CMEs occurred, respectively. The numbers shown indicate the total number of CME events identified in each phase.}
    \label{fig:Sunspot}
\end{figure}

\subsection{Classification of CMEs by Solar Cycle Phase} \label{subsec:class CME}

To study the impact of the solar cycle on CME properties, we classify them using the HELIO4CAST catalog into active phase (AP) events and quiet phase (QP) events based on the solar cycle period they occurred in. Figure~\ref{fig:Sunspot} shows the variation of 13-month smoothed sunspot number (SSN) around the two solar cycle phases. We use a threshold for AP events of ${\geq 70 }$, and of ${\leq 30}$ for QP events, determined visually to capture solar maxima and solar minima from SC23 to SC25. The events that fall in between these two thresholds are called intermediate events, which could potentially exhibit mixed features of both AP and QP events, making it difficult to specifically call them AP or QP events. For this reason, they were not included in this investigation.

\subsection{Heliocentric Distance Binning}

To highlight the effect of heliocentric distance on CME properties, we binned CME events from the HELIO4CAST catalog  based on the heliocentric distance of the spacecraft measuring the CME. For events between 0.3~au and 1~au, we used a binning interval of 0.06~au, and a larger interval of 0.24~au for events spanning 1.0~au to 2.2~au (due to the lower number of events measured at these distances). This approach resulted in a total of 20 bins. For each bin, we perform the coordinate transformation determined using MVA. We then apply the SEA method to magnetic field components in both the RTN and MVA coordinates to examine their variation.

\section{CME Solar Cycle Dependency: Active Phase (AP) vs.\ Quiet Phase (QP) Events}

\subsection{Number and Distribution of Events}
The analysis presented in this section focuses on CMEs observed near $1~au$. Following the CME classification described in Section~\ref{subsec:class CME} for solar cycle dependence, and as illustrated in Figure~\ref{fig:Sunspot}, the distribution of events across the solar cycle phases provides insight into the number of CMEs observed during different levels of solar activity.

\begingroup
\squeezetable
\begin{table}[!hbt]
\centering
\caption{\label{tab:table1} The distribution of CME substructural profiles (magnetic field and plasma parameters) during the active and quiet phases. All values are for measurements near 1 au.} 
\begin{tabular}{lccccc}
\hline
 Sub-region & \multicolumn{3}{c}{AP Event}  &  \multicolumn{2}{c}{QP Event}    \\ 
\cline{2-3} \cline{5-6} 
&  Magnetic Field & Plasma &   &  Magnetic Field & Plasma \\    
\hline
Pre-SW &  428 & 425 &  & 292 & 288 \\       
Sheath &  299 & 294 &  & 189 & 183\\      
ME &  428 & 425 &  & 292 & 290 \\
Wake &  428 & 425 &  & 292 & 288\\
\hline
\end{tabular}
\end{table}
\endgroup

For both solar cycle~23 and~24, the number of events grouped under active phase (AP) is approximately twice the number of quiet phase (QP) events. The selected events used in this study extend only through March 2023, so based on our classification, the number of events per phase in each cycle is as follows: cycle~23 (minimum: 97; maximum: 120), cycle~24 (minimum: 175; maximum: 337), and cycle~25 (minimum: 90; maximum: 60). Also, Table~\ref{tab:table1} presents the distribution of CME substructural profiles for both magnetic field and plasma parameters during the active and quiet phases. The values indicate the number of CME events in each region. About 70\% of CMEs are preceded by a sheath in AP, but only 65\% during QP. Those numbers of CMEs with sheaths are overall consistent with past studies \citep[]{Jian2006,Salman2020}.

\subsection{Average Properties}

Table~\ref{tab:avg_params_Overall} presents the mean values of magnetic and plasma parameters in the pre-SW, sheath, ME, and wake regions, based on the SEA for all events categorized by solar cycle phase. These values represent the average of the mean profile across the time interval corresponding to each substructure. We also report the uncertainties as one standard deviation, capturing both the inherent fluctuations and the systematic temporal evolution within each region.
\begin{table}[ht]
\centering
\caption{Average values from the overall distribution of magnetic and plasma parameters (with one standard deviation of the event distribution) in the pre-SW, sheath, ME, and wake regions for both active (AP) and quiet (QP) CME events. Here, $B_{\text{mean}}$ is the mean magnetic field magnitude, $n_p$ is the proton number density, $V_p$ is the proton bulk speed, $T_p$ is the proton temperature, $M_A$ is the Alfv{\'e}n Mach number, $P_{ram}$ is the solar-wind dynamic (ram) pressure, and $\beta$ is the proton plasma beta.}

\scriptsize
\setlength{\tabcolsep}{4pt}
\begin{tabular}{l@{\hskip 4pt}cc@{\hskip 4pt}cc@{\hskip 4pt}cc@{\hskip 4pt}cc}
\hline
 & \multicolumn{2}{c}{pre-SW} & \multicolumn{2}{c}{sheath} & \multicolumn{2}{c}{ME} & \multicolumn{2}{c}{wake} \\
 & quiet & active & quiet & active & quiet & active & quiet & active \\
\hline
$B_{\text{mean}}$ [nT] & $4.3 \pm 1.8$ & $5.9 \pm 2.6$ & $8.1 \pm 3.9$ & $11.1 \pm 5.3$ & $10.0 \pm 4.5$ & $10.6 \pm 4.8$ & $6.3 \pm 2.3$ & $6.9 \pm 3.3$ \\
$n_p$ [cm$^{-3}$] & $7.1 \pm 4.2$ & $6.2 \pm 4.7$ & $14.4 \pm 8.1$ & $15.3 \pm 10.8$ & $8.5 \pm 4.3$ & $6.5 \pm 5.2$ & $8.1 \pm 4.4$ & $6.0 \pm 4.0$ \\
$V_p$ [km/s] & $367.9 \pm 68.8$ & $398.0 \pm 84.3$ & $412.4 \pm 98.3$ & $485.0 \pm 122.5$ & $403.5 \pm 88.8$ & $462.4 \pm 112.5$ & $408.6 \pm 80.2$ & $449.4 \pm 96.4$ \\
$T_p$ [$\times10^{4}$ K] & $8.0 \pm 6.3$ & $12.3 \pm 14.7$ & $13.5 \pm 19.3$ & $26.6 \pm 25.9$ & $9.9 \pm 11.9$ & $16.3 \pm 16.7$ & $13.6 \pm 10.4$ & $16.9 \pm 14.8$ \\
$M_A$ & $11.5 \pm 4.1$ & $8.0 \pm 2.9$ & $9.8 \pm 3.6$ & $8.4 \pm 3.2$ & $5.6 \pm 1.9$ & $5.1 \pm 1.9$ & $9.3 \pm 3.3$ & $7.7 \pm 3.1$ \\
$P_{ram}$ [$nPa$] & $1.6 \pm 0.9$ & $1.6 \pm 1.3$ & $4.1 \pm 3.3$ & $6.0 \pm 5.1$ & $2.2 \pm 1.2$ & $2.3 \pm 2.1$ & $2.1 \pm 1.1$ & $2.0 \pm 1.4$ \\
$\beta$ & $1.2 \pm 0.9$ & $0.9 \pm 0.9$ & $1.2 \pm 1.0$ & $1.3 \pm 1.3$ & $0.3 \pm 0.3$ & $0.4 \pm 0.4$ & $1.2 \pm 1.0$ & $0.9 \pm 0.9$ \\
\hline
\end{tabular}
\label{tab:avg_params_Overall}
\end{table}

From Table~\ref{tab:avg_params_Overall}, the pre-SW region shows clear AP--QP contrasts. Compared to QP, AP exhibits a higher magnetic field and hotter plasma, with $B_{\mathrm{mean}}$ enhanced by a factor of $1.4\pm0.8$ and $T_p$ by a factor of $1.5\pm2.2$. The upstream density is lower in AP (ratio $0.9\pm0.8$), while the bulk speed is higher by a factor of $1.1\pm0.3$.

The sheath exhibits the largest phase dependence of all substructures, with the most pronounced contrast occurring in the temperature, which increases by a factor of $2.0\pm3.4$ in AP, consistent with stronger sheath heating and compression during AP. In contrast, the sheath density shows only weak phase dependence (ratio $1.1\pm1.0$), despite the strong AP enhancements in the sheath magnetic field strength and ram pressure. 

Inside the ME, $B_{\mathrm{mean}}$, $V_p$, and $T_p$ remain enhanced in AP and the density remains higher in QP, consistent with the trends observed in the other substructures. However, the ME values of $M_A$, $\beta$, and $P_{\mathrm{ram}}$ show only weak phase dependence within the event-to-event variability, suggesting that these quantities are less sensitive to solar cycle phase once inside the ejecta.

We also analyze these properties with Welch's unpaired t-tests on the event-wise mean values in all the substructures (since AP and QP are not matched). In the pre-SW region, the enhancement of $B_{\mathrm{mean}}$ and the reduction of $M_A$ during AP are statistically significant ($\mathit{p} < 0.001$), whereas the differences in $N_p$, $V_p$, $T_p$, and $P_{\mathrm{ram}}$ are not significant ($p > 0.05$), indicating that the upstream phase dependence is primarily magnetic and Alfv{\'e}nic in nature.

Within the sheath, the AP exhibits enhancements in $B_{\mathrm{mean}}$, $T_p$, and $P_{\mathrm{ram}}$  with statistical significance ($p < 0.05$), confirming that stronger sheath compression and heating during AP are intrinsic features. In contrast, the sheath $N_p$, $V_p$, and plasma $\beta$ differences are either marginal or not significant.

Inside the ME, most parameters including $B_{\mathrm{mean}}$, $V_p$, $N_p$, $T_p$, $P_{\mathrm{ram}}$, and $\beta$ do not exhibit strong significance AP--QP differences ($p > 0.05$),  but the $M_A$ during AP is significant ($p < 0.01$) implying that the Alfv{\'e}nic character is phase dependent.

\begin{figure}
    \centering
    \includegraphics[width=\textwidth, height=\textheight, keepaspectratio=true]{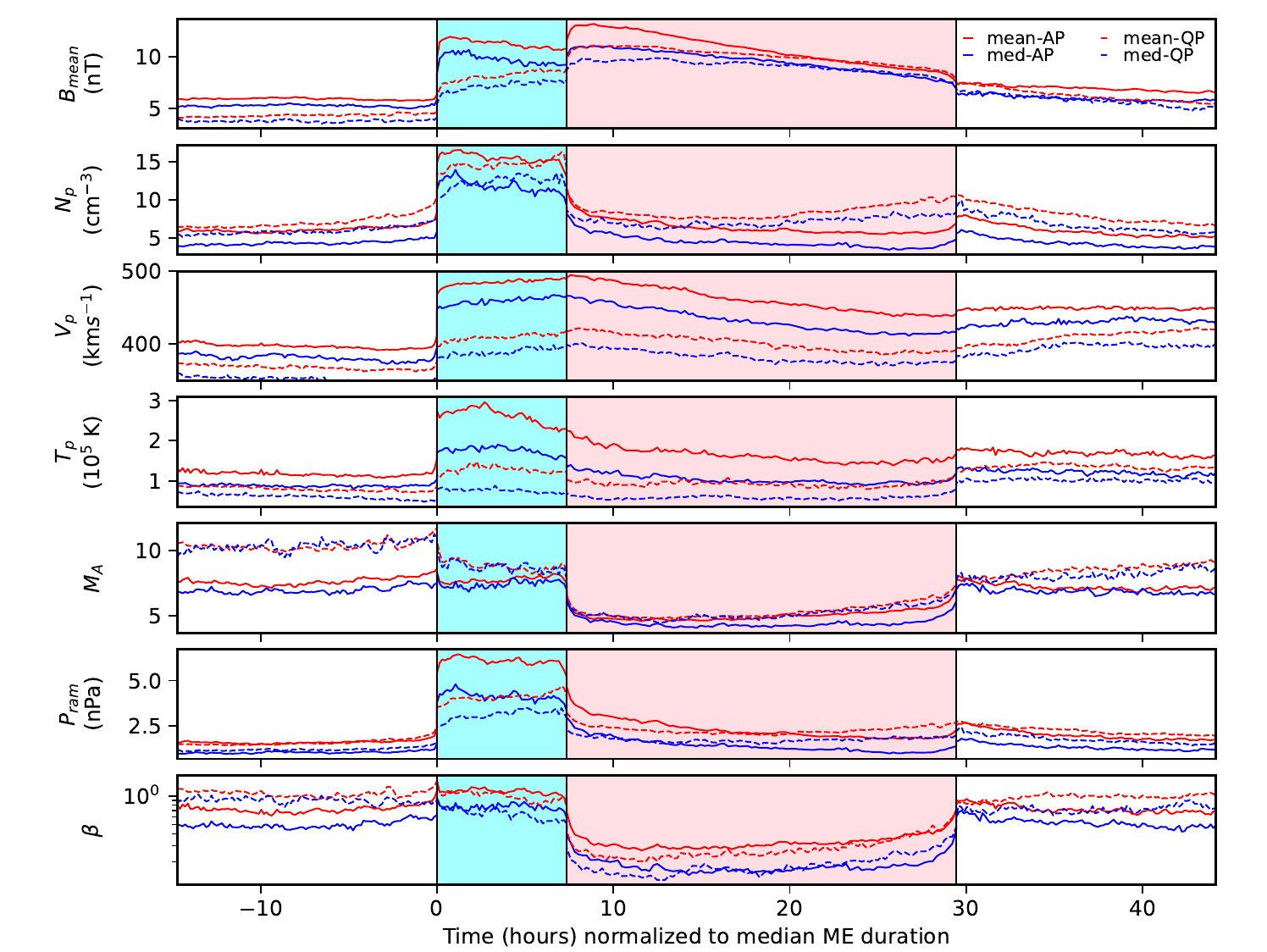}
  \caption{The combined SEA result of AP events (mean in solid red and median in solid blue) and QP events (mean in dashed red and median in dashed blue) profiles, respectively. The cyan area covers the sheath, and the pink area depicts the ME region. From top to bottom, the panels show the magnetic field ($B_{\mathrm{mean}}$), the proton number density ($N_p$), the proton bulk velocity ($V_p$), the proton temperature ($T_p$), the Alfv{\'e}n Mach number ($M_A$), ram pressure ($P_{\mathrm{ram}}$), and proton beta ($\beta$). Here, we used CME events with sheath only.}
    \label{fig:SEA Combine}
\end{figure}

\subsection{SEA on AP and QP Events}\label{sec:Overll SEA}

Figure \ref{fig:SEA Combine} presents the SEA results for the CME events observed by STEREO-A, STEREO-B, and {\it Wind} near 1~au. It also provides a direct comparison between AP and QP event profiles.

First, the magnetic field strength ($B_{\mathrm{mean}}$) is enhanced during AP compared to QP in most regions of the CME except in the ME wake, where both QP and AP have similar magnitudes
The AP profiles display stronger magnetic field peaks in the sheath front, but also shows a gradual decrease of the magnetic field strength towards the rear part of the sheath, unlike the QP profile that  show a gradual rise of the magnetic field strength towards the rear of the sheath. Despite these opposite trends, the sheath remains asymmetric in both phases with a clear enhancement of AP magnetic field near the sheath front and QP at the rear; this may reflect  uneven magnetic compression throughout the sheath region.

At the ME front half, both profiles of AP and QP are enhanced relative to their sheath values. This differs from the findings of \citet{Regnault2020} who reported a decrease in magnetic field strength when transitioning from the sheath into the ME. The origin of this difference is not fully clear from the present comparison and may reflect differences in event categorization, superposed-epoch normalization, or sample composition, rather than a simple sheath-driving versus non-sheath-driving event selection.Following the interpretation of \citep{Siscoe2008}, this enhancement may result from the compression of the sheath-ME interface. Furthermore, the compressed field may drape into the ME front, accumulating and generating a pileup that strengthens the magnetic field in this region. Another possibility is that the ME could be more eroded at its rear due to stronger solar wind interaction or turbulence; however, we cannot yet fully justify this interpretation, as it requires further investigation, which we plan to address in follow-up work. Also, within the ME, the AP profile rapidly declines towards the rear, but the QP retains a smoother and flatter profile throughout the region. 

A front-to-back decrease in field profile across the ME has been linked to ``aging" effects during propagation \citep{Regnault2023b}, and the steeper AP decrease may suggest that such aging signatures are more in the AP. However, such asymmetry is also a known signature of CME expansion during spacecraft crossing, where faster-propagating and strongly expanding CMEs exhibit more pronounced front-to-rear gradients \citep{Janvier2019, Mas-Meza2016}. Given the higher V$_{p}$ observed for AP events in this study, the enhanced asymmetry is more likely associated with stronger expansion dynamics, although a contribution from aging effects cannot be excluded.

 Inside the ME, particularly in the front half, the proton densities (second panel) of both AP and QP events remain ``low'' (comparable to that in the solar wind) and nearly identical. Just before the ME midpoint, the QP N$_p$ gradually increases and becomes enhanced relative to the AP profile. In the rear half of the ME, this separation becomes clearer, with the QP N$_p$ remaining consistently higher than AP up to the ME--Wake boundary.

 This behavior suggests that QP CMEs tend to exhibit more denser plasma in the rear region, probably due to interaction with subsequent high-speed streams or SIRs, whereas AP CMEs might show lower rear densities due to stronger erosion, turbulence, and interaction effects that are more common during solar maximum. The decreasing AP density within the ME is also consistent with expansion and aging. In the wake of the CME, both profiles relax toward the background solar wind ambient level, but with QP remaining slightly elevated.

The AP proton speed ($V_p$, third panel) is higher across all CME substructures, especially within the sheath and ME regions but also in the pre-existing solar wind and wake; the signature tends to decrease from the front to the back of the ME, which is an indication of radial expansion, as reported by \citet{Klein1982} and \citet{Farrugia1993}. This higher speed in AP enhances shock formation and sheath compression. QP events, on the other hand, display a reduced velocity profile, with a much smaller jump at the start of the sheath,  which suggests a less energetic ejection propagation process. We find that AP events radially expand with an average speed of 22.5~km~s$^{-1}$, compared to 13.6~km~s$^{-1}$ for QP events. This SEA-based expansion speed is computed as half the difference between the front and rear velocities of the magnetic ejecta, $V_{\mathrm{exp}}=(V_{\mathrm{front}}-V_{\mathrm{rear}})/2$. As a consistency check, we also estimate expansion speeds on an event-by-event basis using the same definition applied to each individual CME velocity profile, and then average across events. This yields mean expansion speeds of 23.6~km~s$^{-1}$ for AP and 13.3~km~s$^{-1}$ for QP, which are consistent with the SEA-derived estimates.

The AP CMEs exhibit an increased temperature profile ($T_p$, fourth panel) over all the regions (especially in the sheath), indicating stronger plasma heating due to their higher speed, as shown in the third panel. QP CMEs, in contrast, maintain a lower temperature profile, which is consistent with reduced interaction and energy during propagation. This difference in thermal behavior aligns with \citep{Mas-Meza2016, Richardson1995}, who showed that faster CMEs originating from active periods generally exhibit higher temperatures. Additionally, the higher temperatures observed for AP events may also reflect reduced cooling time associated with faster propagation, as well as enhanced compression and density effects, particularly toward the rear of the ME where interactions with high-speed streams can increase both density and temperature.

The last three panels present the Alfv{\'e}n Mach number ($M_A$), ram pressure ($P_{\mathrm{ram}}$), and proton plasma beta ($\beta$), respectively, as temporal profiles of the CMEs. The ram pressure directly depends on the flow speed and plasma density, therefore, an increase in density or flow speed produces a stronger dynamic force on the magnetosphere, which may enhance magnetopause compression. The Alfv\'en Mach number also depends on the flow speed and density but also on the magnetic field. Inside the ME, the Alfv\'en Mach number is low, which has been shown to affect the solar wind-magnetosphere coupling \citep[]{Farrugia1995, Lavraud2008}. The proton plasma beta, which is the ratio  of the thermal pressure and magnetic pressure, is also influenced by similar variations, i.e., higher density and temperature tend to increase $\beta$, while stronger magnetic field structures reduce it. The effect of magnetic forces on solar wind flow in the magnetosheath scales approximately as $1/M_{A}^{2}$, meaning that magnetic control is always stronger when $M_{A}$ is low, but weaker when $M_{A}$ is high. 

Across all regions (especially within the sheath and wake region), the quiet phase (QP) events exhibit higher Alfv{\'e}n Mach numbers ($M_A$) compared to the active phase (AP) events, indicating that the CME-driven solar wind during QP is generally more super-Alfv{\'e}nic and therefore less influenced by magnetic forces in the magnetosheath. In contrast, the lower $M_A$ values observed during AP in these regions indicate that the solar wind may be closer to the Alfv{\'e}nic regime, and this may enhance the role of magnetic tension and pressure in controlling the plasma dynamics. However, within the ME, the $M_A$ profiles show no significant difference between AP and QP, with averaged values of $\langle M_A \rangle_{\mathrm{ME}} = 5.6$ (AP) and $5.1$ (QP). This suggests that the solar wind magnetosphere coupling during the ME passage may be similar in both phases. This AP trend in $M_A$ is also consistent with the observed reduction in proton plasma beta ($\beta$) during AP, particularly within the ME, where the low-$\beta$ state suggests a more magnetically dominated structure during the active solar phase. Additionally, the ram pressure profile shows that AP events have significantly greater dynamic pressure in the sheath and ME front, while QP events exceed AP toward the rear of the ME, so this may suggest differences in CME expansion and interaction with the background solar wind across the solar cycle.

Overall, the SEA profiles suggest enhanced magnetic field strength, temperature, and propagation speed for AP CMEs compared to QP events, while QP events tend to exhibit higher plasma densities. However, statistical T-tests indicate that the differences in magnetic field strength, temperature, and speed are not significant, whereas the higher plasma density observed in QP events is statistically supported.

These trends nevertheless point to a possible influence of solar cycle phase on CME evolution. They may also have implications for space weather forecasting, as CMEs during AP could potentially lead to stronger geomagnetic disturbances.

\subsection{Difference in CME Properties Controlling for the ME Speed: SEA}

Here, we aim to control the potential influence of the CME speed on how solar cycle activity impacts CME properties by creating a subset of the AP events with a comparable ME velocity distribution as the QP events. The goal is to determine whether or not the differences between AP and QP events as described in section~\ref{sec:Overll SEA} are due predominantly to the speed difference between the events. It is for example well-known that CMEs with higher velocities tend to have higher magnetic field strength \citep[]{Moestl2014}. We create a subset of both the AP and QP CMEs which have similar speed properties (average and standard deviation) by applying a one-to-one nearest-neighbor matching based on the event-wise mean ME speed, with a maximum allowed difference of 10~km~s$^{-1}$. After applying the speed matching criterion, we were left with 133 events for both AP and QP, respectively. The resulting ME speed distributions are nearly identical, with median speeds of 404.8 km\,s$^{-1}$ (standard deviation 78.8 km\,s$^{-1}$) for AP and 405.6 km\,s$^{-1}$ (standard deviation 77.8 km\,s$^{-1}$) for QP.

\begin{table}[ht]
\centering
\caption{Average values from the distribution of magnetic and plasma parameters (with one standard deviation of the event distribution) in the pre-SW, sheath, ME, and wake regions for a subset of both active (AP) and quiet (QP) CME events for which the speed is matched.}
\scriptsize
\setlength{\tabcolsep}{4pt}
\begin{tabular}{l@{\hskip 4pt}cc@{\hskip 4pt}cc@{\hskip 4pt}cc@{\hskip 4pt}cc}
\hline
 & \multicolumn{2}{c}{pre-SW} & \multicolumn{2}{c}{sheath} & \multicolumn{2}{c}{ME} & \multicolumn{2}{c}{wake} \\
 & quiet & active & quiet & active & quiet & active & quiet & active \\
\hline
$B_{\text{mean}}$ [nT] & $4.3 \pm 1.8$ & $5.3 \pm 2.2$ & $8.1 \pm 3.6$ & $9.4 \pm 4.2$ & $9.8 \pm 3.8$ & $10.3 \pm 4.3$ & $6.1 \pm 2.2$ & $6.8 \pm 3.1$ \\
$n_p$ [cm$^{-3}$] & $7.2 \pm 4.4$ & $6.7 \pm 5.6$ & $13.7 \pm 7.3$ & $16.0 \pm 11.2$ & $8.3 \pm 4.3$ & $7.3 \pm 5.5$ & $7.8 \pm 4.5$ & $6.8 \pm 3.9$ \\
$V_p$ [km/s] & $370.3 \pm 67.7$ & $366.2 \pm 66.7$ & $415.3 \pm 89.8$ & $425.1 \pm 93.8$ & $407.4 \pm 79.0$ & $407.7 \pm 79.5$ & $410.2 \pm 71.6$ & $411.0 \pm 76.6$ \\
$T_p$ [$\times10^{4}$ K] & $8.4 \pm 6.3$ & $9.1 \pm 6.6$ & $13.1 \pm 16.5$ & $17.0 \pm 17.8$ & $9.7 \pm 8.7$ & $11.2 \pm 8.9$ & $13.5 \pm 9.1$ & $13.9 \pm 10.5$ \\
$M_A$ & $11.7 \pm 4.2$ & $8.6 \pm 3.5$ & $9.6 \pm 3.1$ & $8.8 \pm 3.4$ & $5.7 \pm 2.0$ & $4.9 \pm 1.7$ & $9.3 \pm 3.2$ & $8.1 \pm 3.5$ \\
$P_{ram}$ [$nPa$] & $1.6 \pm 1.0$ & $1.5 \pm 1.3$ & $3.9 \pm 3.0$ & $4.7 \pm 3.6$ & $2.2 \pm 1.2$ & $2.0 \pm 1.7$ & $2.1 \pm 1.2$ & $1.9 \pm 1.2$ \\
$\beta$ & $1.3 \pm 1.0$ & $1.0 \pm 1.1$ & $1.1 \pm 0.9$ & $1.3 \pm 1.4$ & $0.3 \pm 0.3$ & $0.3 \pm 0.3$ & $1.2 \pm 1.0$ & $0.9 \pm 0.8$ \\
\hline
\end{tabular}
\label{tab:Velocity-matching}
\end{table}

Figure~\ref{fig:SEA_Matiching Combine} presents the resulting temporal profiles after velocity matching. A region by region comparison of Figures~\ref{fig:SEA Combine} and~\ref{fig:SEA_Matiching Combine} shows that the temporal morphology of the $B_{\mathrm{mean}}$ profile remains largely unchanged across all CME substructures, from the pre-SW through the sheath, ME, and wake, although the magnitudes differ between the two cases. In both figures, AP events exhibit enhanced magnetic fields relative to QP, and the overall field evolution (including the ME front side enhancement relative to the entire sheath and the subsequent decline toward the wake) is preserved. This morphological stability suggests that the solar cycle dependence of the magnetic field profile is not impacted by controlling the ME $V_p$, and that the AP--QP differences in $B_{\mathrm{mean}}$ profiles observed in Figures~\ref{fig:SEA Combine} and \ref{fig:SEA_Matiching Combine} cannot be attributed solely to differences in CME speed but may likely reflect intrinsic differences in CME magnetic structure across solar cycle phases.

However, there is a notable change in the $N_p$ and $V_p$ profiles within the ME and the pre-SW following the region by region comparison between Figures~\ref{fig:SEA Combine} and \ref{fig:SEA_Matiching Combine}. For $N_p$, the AP and QP profiles become nearly identical in the front side of the pre-SW region and in the earlier portion of the ME for Figure~\ref{fig:SEA_Matiching Combine}, whereas the sheath and wake $N_p$ profiles remain unchanged in both Figures~\ref{fig:SEA Combine} and \ref{fig:SEA_Matiching Combine}. Likewise, for $V_p$, the AP--QP separation in the pre-SW and within the early portion of ME that is evident in Figure~\ref{fig:SEA Combine} is quite reduced in Figure~\ref{fig:SEA_Matiching Combine}, just downstream of the ME front and prior to the midpoint. However, the sheath $V_p$ profiles remain distinct between AP and QP in both figures. The decrease in the AP--QP separation within the ME suggests that some of the ME differences observed in Section~\ref{sec:Overll SEA}, especially those linked to the ME front $N_p$ and the early ME $V_p$ evolution, were at least partly speed mediated.

The $T_p$, $M_A$, $P_{\mathrm{ram}}$, and $\beta$ profiles remain consistent between Figure~\ref{fig:SEA_Matiching Combine} and Figure~\ref{fig:SEA Combine} across all substructures. This persistence, together with the continued phase dependence of sheath $V_p$ even after ME speed matching, indicates that the solar cycle dependence identified in Section~\ref{sec:Overll SEA} is not primarily an artifact of CME speed differences. While controlling for ME speed reduces some AP--QP contrasts in $N_p$ and $V_p$ near the ME front, the differences in $T_p$, $M_A$, $P_{\mathrm{ram}}$, and $\beta$ persist in both analyses, particularly within the sheath. This suggests that ambient solar cycle conditions and CME--solar-wind interaction processes contribute to the observed phase dependence beyond the effect of CME speed alone.

We perform paired t-tests on the event-wise mean values for each parameter and region to assess the significance of AP--QP differences in the speed-matched distribution. In the ME region, the bulk speed difference is not significant ($p = 0.27$), confirming that the speed-matching procedure successfully removes systematic phase-dependent speed differences. Likewise, the ME $B_{\mathrm{mean}}$, $N_p$, $T_p$, $P_{\mathrm{ram}}$, and plasma $\beta$ show no significant AP--QP differences ($p > 0.05$), indicating that most core plasma differences are largely controlled by CME speed.

In contrast, within the sheath region, CME properties are significantly enhanced during AP for $B_{\mathrm{mean}}$ (p = $8.7\times10^{-4}$), $T_p$ ($p = 0.025$), and $P_{\mathrm{ram}}$ ($p = 0.032$). This suggests that the stronger sheath compression observed during AP is not solely driven by CME speed differences. This contrasting pattern of consistency between phases reinforces the presence of solar-cycle-dependent structuring.

\subsection{Difference in CME Properties Controlling for the ME Speed: Average Properties}

\begin{figure}
    \centering
    \includegraphics[width=\textwidth, height=\textheight, keepaspectratio=true]{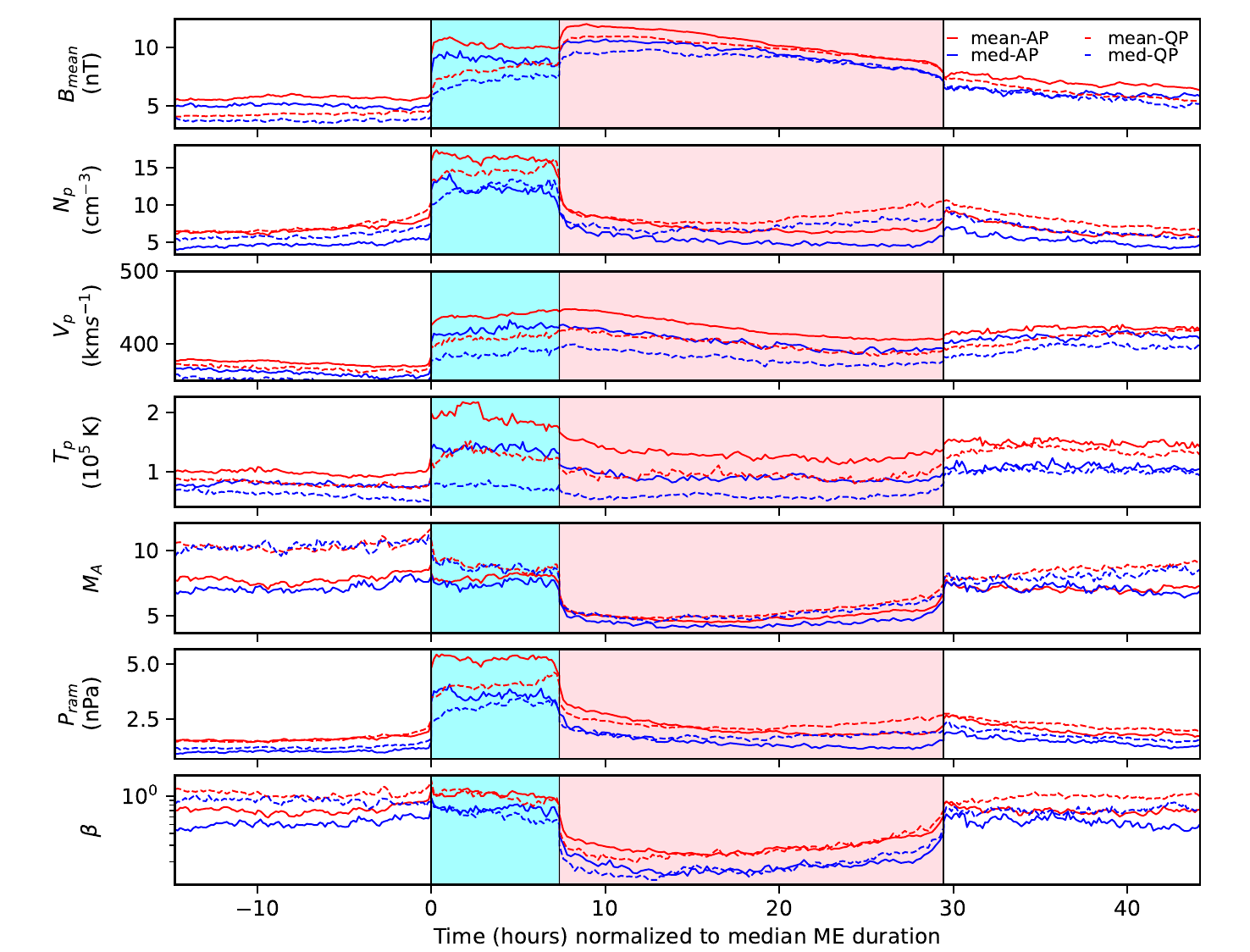}
    \caption{The combined SEA result of AP events (mean in solid red and median in solid blue) and QP events (mean in dashed red and median in dashed blue) profiles, respectively. Parameters are displayed in the same format as Figure~2.}
    \label{fig:SEA_Matiching Combine}
\end{figure}

We next examine the profile-based results shown in Figure~\ref{fig:SEA_Matiching Combine} from a statistical perspective using Table~\ref{tab:Velocity-matching}. Table~\ref{tab:Velocity-matching} summarizes the region-averaged magnetic and plasma parameters (mean $\pm$ one standard deviation of the event distribution) in the pre-SW, sheath, ME, and wake regions for the ME speed-matched AP and QP events. The ME bulk speed becomes nearly identical between the two phases, with $\langle V_p\rangle_{\mathrm{ME}} = 407.7 \pm 79.5$~kms$^{-1}$ for AP and $407.4 \pm 79.0$~kms$^{-1}$ for QP, confirming that the speed-matching procedure controls the ME speed. Consistent with the reduced separation in the $V_p$ profiles within the ME in Figure~\ref{fig:SEA_Matiching Combine}, the pre-SW and wake speeds also become comparable between phases. In contrast, the sheath remains faster in AP than in QP ($425.1 \pm 93.8$~kms$^{-1}$ versus $415.3 \pm 89.8$~kms$^{-1}$), indicating that sheath flow differences persist even after ME speed matching.

Despite the controlled ME speed, solar cycle dependent differences remain apparent in some CME properties considered. In particular, the magnetic field strength remains higher in AP than in QP across all substructures, including within the speed-matched ME region with values $\langle B_{mag}\rangle_{\mathrm{ME}} = 10.3 \pm 4.3$~nT for AP versus $9.8 \pm 3.8$~nT for QP. This is consistent with the SEA profiles obtained in section~\ref{sec:Overll SEA}, whose ME-averaged $B_{\mathrm{mean}}$ values are ($10.6 \pm 4.8$~nT for AP and $10.0 \pm 4.5$~nT for QP, see Table~\ref{tab:avg_params_Overall}), and indicates that the AP--QP difference in $B_{\mathrm{mean}}$ is not driven only by CME speed (Figure~\ref{fig:SEA_Matiching Combine}). The sheath compression signatures also persist statistically, with higher sheath $N_p$ ($16.0 \pm 11.2$~cm$^{-3}$ versus $13.7 \pm 7.3$~cm$^{-3})$ and higher sheath ram pressure ($4.7 \pm 3.6$~nPa versus $3.9 \pm 3.0$~nPa) during AP, consistent with the sustained AP enhancement observed in the sheath SEA profiles.

Other plasma properties also remain consistent with their SEA profile behavior. For example, the ME exhibits low proton plasma $\beta$ in both AP and QP, with nearly identical values of $\langle\beta\rangle_{\mathrm{ME}} = 0.3 \pm 0.3$ (AP) and $0.3 \pm 0.3$ (QP), as well as low Alfv{\'e}n Mach numbers in both phases. 
Taken together, Table~\ref{tab:Velocity-matching} reinforces the interpretation drawn from Figures~\ref{fig:SEA Combine} and~\ref{fig:SEA_Matiching Combine}: although controlling the ME speed reduces AP--QP differences in ME region for $V_p$ and $N_p$ profiles, the persistence of stronger AP magnetic fields and enhanced sheath compression indicates that the solar cycle dependence of CME properties near 1~au is not primarily driven by speed alone.

\section{CME Heliocentric Distances Dependency}
\subsection{SEA of CMEs at Different Heliocentric Distances}

\begin{figure}[hbt]
    \centering
    \includegraphics[width=0.9\textwidth]{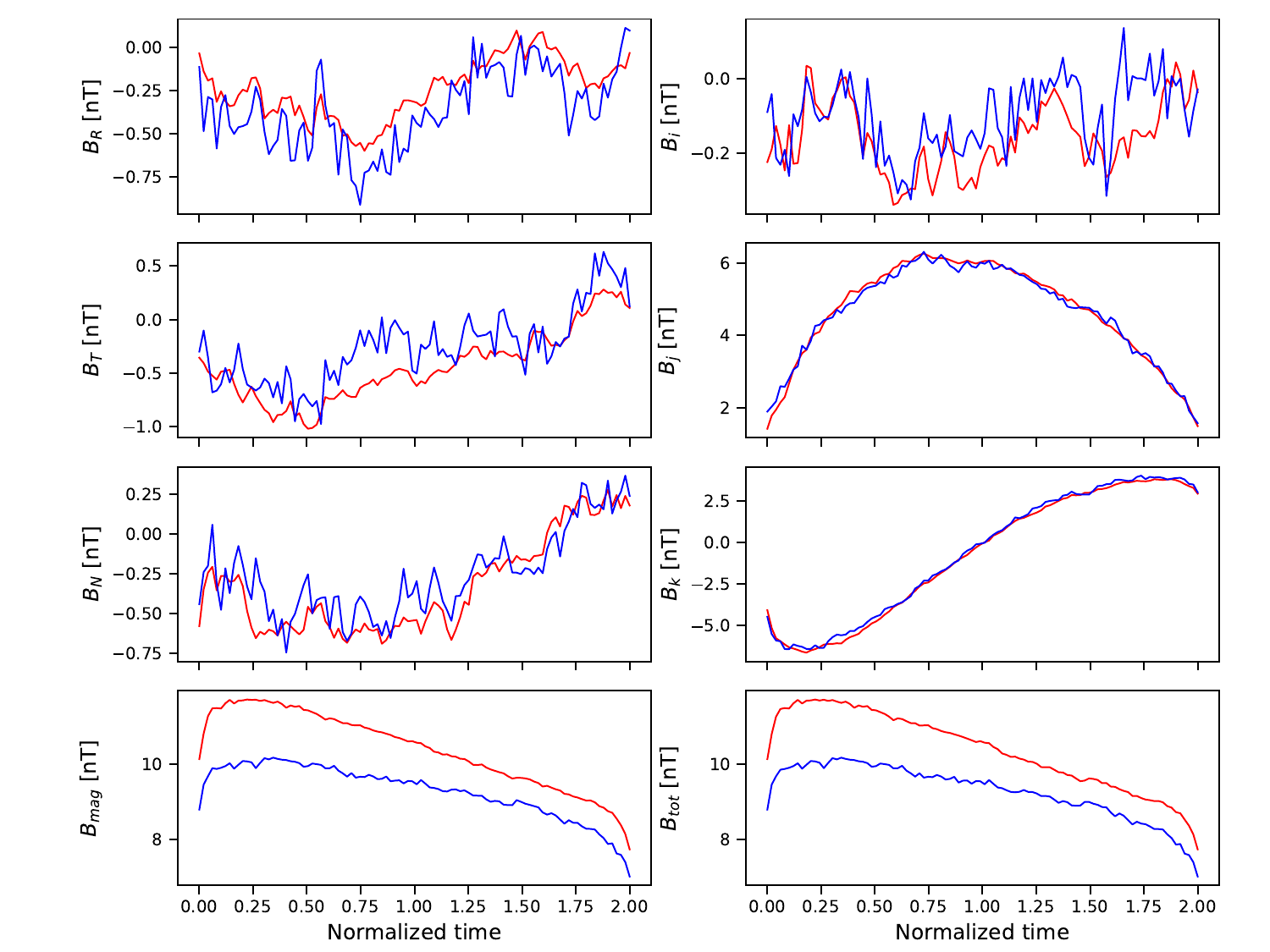}
    \caption{Sample SEA for magnetic field components (red for mean profiles and blue for median profile) in RTN vs.\ MVA coordinates for CMEs in Bin~17 (0.96~au to 1.02~au). On the left, panels 1-3 show the results in RTN components, and panel 4 shows the total field strength. The right plot is for MVA coordinate components (ijk), whose magnitude B$_{tot}$ is displayed in panel 4.}
    \label{fig:Bin17_SEA}
\end{figure}

We next turn our attention to the SEA profile of CMEs at different heliocentric distances. To include the CME magnetic field components, we perform minimum variance analysis (MVA) and present the result in the $i,j,k$ MVA frame \citep{Lanabere2022}.
We group all CME events into 20 bins based on their heliocentric distance, as summarized in Table~\ref{tab:heliocentri}, and perform SEA on each bin to examine the variation of their mean temporal profiles. As an example, the SEA result for Bin~17 is shown in Figure~\ref{fig:Bin17_SEA}. Table~\ref{tab:heliocentri} also reports the distance range of each bin, the number of events before and after MVA (due to eliminating events that do not match the criterion for the ratio of eigenvalues), the maximum $B_{k}$ front-to-rear ratio, and the mean and median magnetic field magnitudes ($\langle B_{\mathrm{mean}}\rangle$ and $\langle B_{\mathrm{median}}\rangle$), along with the $\langle |B_{k}|_{\mathrm{mean}}\rangle$ and $\langle |B_{j}|_{\mathrm{mean}}\rangle$. 
Bins 1--4 are excluded due to a lack of events, as the ICMECAT v2.1 catalog used predates the PSP exploration of these heliocentric distances.


\begingroup
\squeezetable
\begin{table}
\centering
\caption{\label{tab:heliocentri} All event bins grouped by heliocentric distance intervals. Bins 1--4 (0.00--0.24 au) are omitted because no events fall within these ranges. Bins marked with an asterisk contain events from the Juno mission (1.00--2.20 au); for these bins, a mid-bin value of 0.24 au is adopted, while the remaining bins use 0.06 au spacing.}
\setlength{\tabcolsep}{4pt}
\small
\begin{tabular}{lccccccccc}
\hline
Bin & Mid-bin & $r_{\mathrm{range}}$ & Event & Event & $B_{k\text{-front}}/$ & $\langle B_{\mathrm{mean}} \rangle$ & $\langle B_{\mathrm{med}} \rangle$ & $\langle |B_{k}|_{\mathrm{mean}}\rangle$ & $\langle |B_{j}|_{\mathrm{mean}}\rangle$ \\
 &  & (au) & (\#) & $_{\mathrm{MVA}}$ & $B_{k\text{-rear}}$ & (nT) & (nT) & (nT) & (nT) \\
\hline
\hline
5  & 0.27 & 0.24-0.30 & 3   & 3   & 1.02 & 74.4 & 74.8 & 45.3 & 35.1 \\ 
6  & 0.33 & 0.30-0.36 & 34  & 28  & 1.42 & 49.7 & 49.4 & 25.6 & 32.0 \\
7  & 0.39 & 0.36-0.42 & 19  & 12  & 1.18 & 42.3 & 41.5 & 20.4 & 28.8 \\
8  & 0.45 & 0.42-0.48 & 37  & 33  & 1.68 & 42.7 & 43.4 & 23.3 & 27.0 \\
9  & 0.51 & 0.48-0.54 & 7   & 6   & 0.74 & 22.7 & 23.0 & 12.2 & 10.4 \\
10 & 0.57 & 0.54-0.60 & 9   & 8   & 1.30 & 21.2 & 21.2 & 12.5 & 10.1 \\
11 & 0.63 & 0.60-0.66 & 8   & 8   & 1.23 & 15.7 & 15.8 & 10.9 & 7.3 \\
12 & 0.69 & 0.66-0.72 & 14  & 14  & 1.24 & 17.7 & 17.9 & 11.5 & 9.2 \\
13 & 0.75 & 0.72-0.78 & 13  & 10  & 1.42 & 18.8 & 18.8 & 10.7 & 11.2 \\
14 & 0.81 & 0.78-0.84 & 11  & 9   & 1.56 & 13.9 & 14.2 & 8.6  & 7.0 \\
15 & 0.87 & 0.84-0.90 & 5   & 5   & 1.66 & 13.3 & 13.6 & 8.4  & 6.3 \\
16 & 0.93 & 0.90-0.96 & 117 & 124 & 1.46 & 9.8  & 9.7  & 5.4  & 5.6 \\
17 & 0.99 & 0.96-1.02 & 667 & 667 & 1.44 & 10.4 & 10.3 & 6.0  & 5.8 \\
18 & 1.05 & 1.02-1.08 & 80  & 78  & 1.79 & 9.0  & 8.8  & 5.1  & 4.9 \\
19 & 1.11 & 1.08-1.14 & 31  & 31  & 1.29 & 8.7  & 8.7  & 5.1  & 4.9 \\
20 & 1.17 & 1.14-1.20 & 4   & 4   & 1.46 & 9.4  & 9.3  & 3.7  & 6.2 \\
21$^{*}$ & 1.36 & 1.24-1.48 & 15  & 15  & 1.42 & 5.6  & 5.7  & 2.9  & 3.0 \\
22$^{*}$ & 1.60 & 1.48-1.72 & 4   & 4   & 1.20 & 4.5  & 4.6  & 3.1  & 2.2 \\
23$^{*}$ & 1.84 & 1.72-1.96 & 9   & 9   & 1.66 & 4.5  & 4.5  & 2.6  & 2.5 \\
24$^{*}$ & 2.08 & 1.96-2.20 & 10  & 10  & 1.39 & 3.4  & 3.4  & 1.9  & 1.7 \\
\hline
\hline
\end{tabular}
\end{table}
\endgroup

The SEA result for Bin~17 (Figure \ref{fig:Bin17_SEA}) shows magnetic field components in both RTN (left panels) and MVA (right panels) coordinate systems. The MVA results reveal a clear rotation between the poloidal field component ($B_j$) and the axial or toroidal field component ($B_k$). To ensure consistent rotation patterns across all bins, we flip the signs of $B_k$ and $B_j$ where necessary. This adjustment is valid because eigenvector directions in MVA are inherently sign-ambiguous, allowing us to maintain a uniform interpretation of field rotation without altering the physical structure of the events.

\subsection{Decrease of the CME Magnetic Field Magnitude With Distance}
\begin{figure}
    \centering
    \begin{minipage}[t]{0.48\linewidth}
        \centering
        \includegraphics[width=\linewidth]{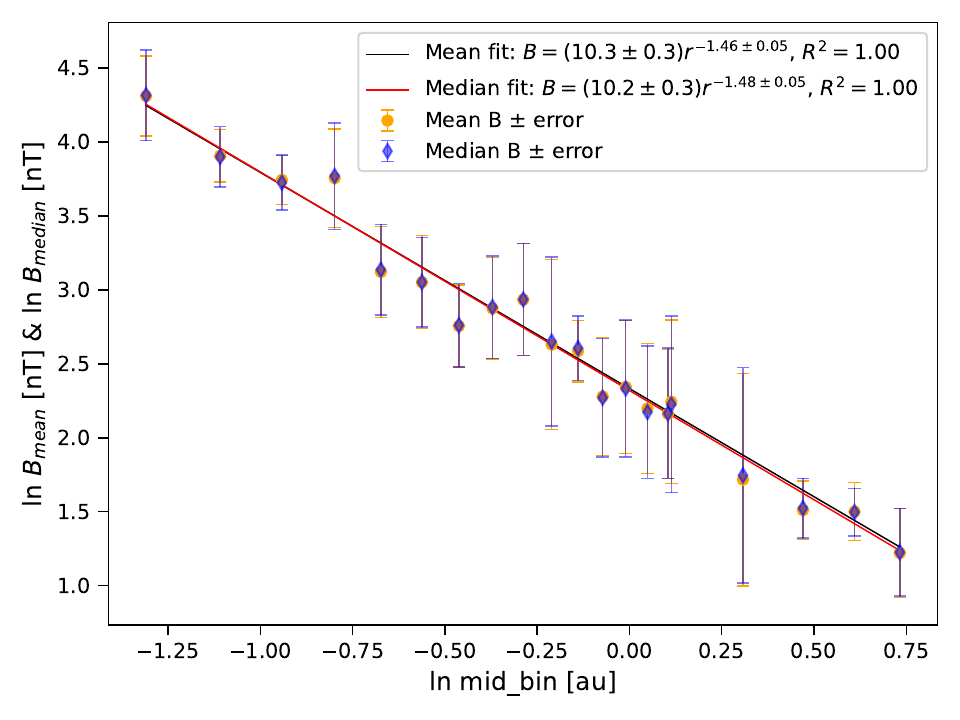}
        
    \end{minipage}
    \hfill
    \begin{minipage}[t]{0.48\linewidth}
        \centering
        \includegraphics[width=\linewidth]{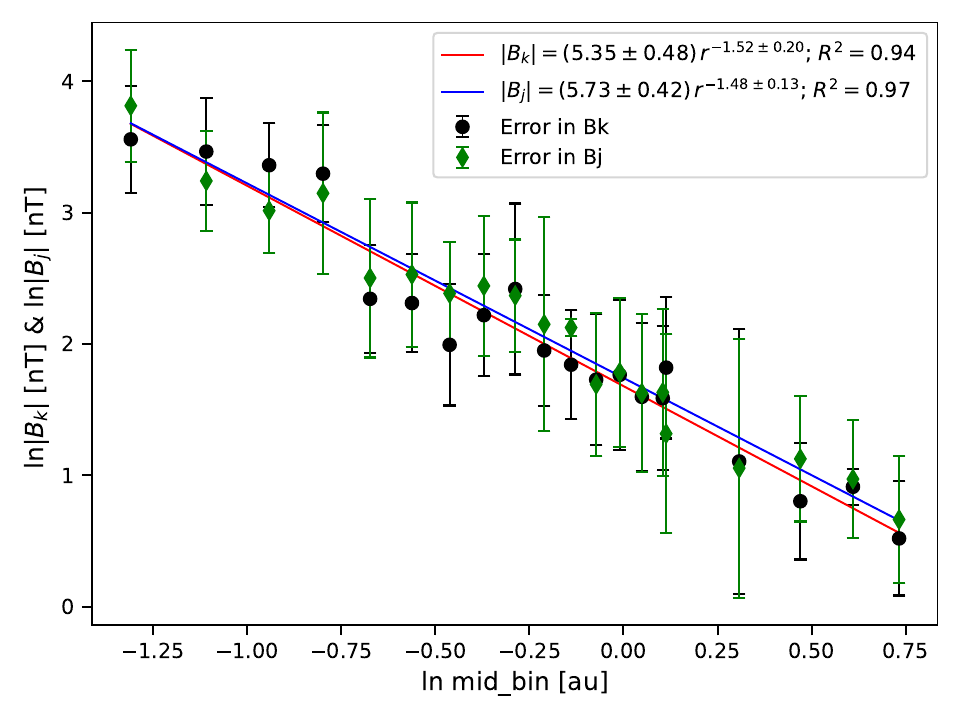}
    \end{minipage}

    \caption{Left: Radial evolution of the average CME magnetic field strength as a function of the natural logarithm of heliocentric distance, shown using both the binned mean (orange circles) and median (blue diamonds) values. The error bars correspond to the $1\sigma$ dispersion of the event distribution within each heliocentric distance bin, converted to logarithmic space using $\Delta \ln B \approx \sigma_B/B$. Solid lines (black for mean and red for median) show robust power-law fits of the form $B=a\,r^{b}$ obtained by fitting $\ln(B)$ versus $\ln(r)$ using a RANSAC linear regressor. The fitted coefficients, exponents, and the corresponding $R^{2}$ values (computed using RANSAC inliers only) are reported in the legend.
    Right: Robust power-law fits for the toroidal ($|B_k|$) and poloidal magnetic field components ($|B_j|$), as functions of heliocentric distance in natural log space. The error bars represent the uncertainty in $\ln(|B|)$ propagated from the standard deviation of each component, $\Delta\ln(|B|)\approx\sigma_{|B|}/|B|$. The fits ($B=a,r^{b}$) are obtained using  RANSAC fitting (red line for $|B_k|$ and blue line for $|B_j|$); the fit equations and corresponding $R^{2}$ values are reported in the legend.} 
    \label{fig:PowerLaw_and_Ransac_plots}
\end{figure}

We next examine how CME properties depend on distance; to do so, we use a Random Sample Consensus (RANSAC) fitting technique in logarithmic space to fit a power law to our dataset plots \citep{Fischler1981}. In this approach, iterative random sampling and consensus evaluation with a deterministic inlier threshold ensure that the model is estimated solely from inliers, unlike ordinary least squares fitting, which is sensitive to outliers. In Figure~\ref{fig:PowerLaw_and_Ransac_plots} (left panel), we present the mean and median magnetic field strengths over different distances in a natural logarithm--natural logarithm (ln--ln) plot. The best fit of the power law relation for our mean magnetic field data is $B_{mean} = (10.3 \pm 0.3)\, r^{-1.46 \pm 0.04}$, and for the median profile, $B_{median} = (10.2 \pm 0.30)\, r^{-1.48 \pm 0.04}$, where $B$ is in nanotesla (nT) and $r$ is the heliocentric distance in astronomical units (au). These fits are derived from the averaged events within each bin. Here, the reported uncertainties are 95\% confidence intervals obtained through bootstrapping analysis, and the symmetric uncertainties in the power law prefactor arise from propagation of the log space fitting errors through the exponential transformation.

Although the power law relations reported by  \citet{Leitner2007}, \citet{Davies2021}, \citet{Christian2026}, and \citet{Wang2005} were derived using datasets spanning broader heliocentric distance ranges than considered here, their reported decay exponents ($-1.64$, $-1.64$, $-1.57$ and, $-1.52$ respectively) are of comparable magnitude to our fitted value ($b=-1.46$). Also, studies that focused on shorter heliocentric ranges near 1~au \citep[e.g.,][]{Mariani1990,Totten1995} reported  exponents of $-1.56$ and $-1.64$ also consistent with our result  but \citet{Winslow2015} and \citet{Gulisano2010}, which were limited to distances up to $\sim$1~au,  reported steeper decays ($-1.95$ and $-1.85$) which indicate more rapid radial weakening than found here.

Among these studies, the reported decay exponents of \citet{Wang2005} and \citet{Mariani1990} are the closest to our binned estimates for both the mean and median profiles. The slightly steeper decay rates reported by \citet{Leitner2007} and \citet{Davies2021} likely reflect methodological differences, including the use of fewer spacecraft differences in event selection, and the use of event-based fitting approaches without explicit heliocentric distance binning. The shared feature of these earlier studies is that the power law fits were typically performed on non-binned datasets spanning their respective distance ranges, rather than within discrete heliocentric distance bins as adopted in this work. To assess the impact of this methodological choice, we also performed a fit on our full dataset without distance binning them and obtained $B_{\mathrm{mean}}=(9.2\pm0.1)\,r^{-1.75\pm0.06}$ and $B_{\mathrm{median}}=(9.1\pm0.1)\,r^{-1.61\pm0.07}$, with exponents that fall within the range reported in previous works. This highlights the fact that, for the same dataset, the resulting power-law can vary significantly based on whether or not the data has been binned. The difference in the power-law index between the binned and non-binned version is significantly higher than the uncertainty given by the RANSAC fit.

Turning to studies that applied heliocentric distance binning, \citet{Hanneson2020} reported decay indices of $r^{-1.51}$ and $r^{-1.47}$ for the power law fitting of the cruise- and orbital-phase binned datasets of the IMF taken by MESSENGER, VEX, STEREO, and ACE, respectively. These values are in close agreement with our results, with their orbital-phase exponent lying between our mean and median estimated exponents. This indicates that the decrease of the CME magnetic field with distance may be in fact very similar to that of the IMF. From our analysis, together with previous studies, it is evident that binned exponents generally indicate a slower radial decay than those derived from non-binned fits. 

\subsection{Decrease of the CME Magnetic Field Components in MVA Frame with Distance} 
Several studies have investigated CME expansion through the analysis of the heliocentric dependence of the individual magnetic field components; for instance, when both the axial and azimuthal components decrease at comparable rates with heliocentric distance \citep[]{Lugaz2020}, the process is described as isotropic expansion in \citet{Yu2024}. \citet{Shimazu2002} and \citet{Berdichevsky2003}, among others, proposed models to describe such isotropic CME expansion.

We now present the analysis of the variation of the MVA magnetic field components with heliocentric distance, which has not been done in the majority of past statistical studies. Figure~\ref{fig:PowerLaw_and_Ransac_plots} (right panel) shows the heliocentric-distance variation of the event-wise mean absolute toroidal ($|B_k|$ or $|B_z|$) and poloidal ($|B_j|$ or $|B_\phi|$) magnetic-field components, averaged within each distance bin, together with their corresponding RANSAC power law fits. The fitted relations, $|B_{k}|\,(\mathrm{nT}) = (5.4 \pm 0.5)\, r^{1.52 \pm 0.20}$ and $|B_{j}|\,(\mathrm{nT}) = (5.7 \pm 0.4)\, r^{1.48 \pm 0.13}$ indicate similar expansion behavior, as reflected in their comparable decay exponents and their magnitudes at 1~au is also similar. Our fitted exponents of $-1.52$ for $|B_{k}|$ and $-1.48$ for $|B_{j}|$ are within approximately 10\% of the values reported by \citet{Yu2024} and \citet{Leitner2007}, who both obtained an index of $-1.64$ based on significantly fewer events. This consistency suggests that both $|B_{k}|$ and $|B_{j}|$ exhibit self-similar expansion with comparable scaling, although \citet{Yu2024} reported non-self-similar expansion behavior between the two components. Some self-similarly expanding flux-rope models predict different radial scalings for the axial and azimuthal components, with $B_z \propto r^{-2}$ and $B_\phi \propto r^{-1}$ \citep[e.g.,][]{Osherovich1993, Farrugia1993}, but our results imply that the differential radial scaling predicted by some idealized models may not fully describe the average large scale evolution of CMEs in the heliosphere. To the best of our knowledge, this work represents one of the first statistical investigations of the axial and poloidal fields evolution inside MEs, which demonstrates that the toroidal and poloidal components weaken concurrently and at comparable rates during propagation. Any theoretical CME model should explain this similar expansion of toroidal and poloidal components.

\subsection{Evolution with Distance of the Poloidal-to-Toroidal and Front-to-Rear Magnetic Field Ratio}

\begin{figure}
    \centering

    \begin{minipage}[b]{0.48\linewidth}
        \centering
        \includegraphics[width=\linewidth]{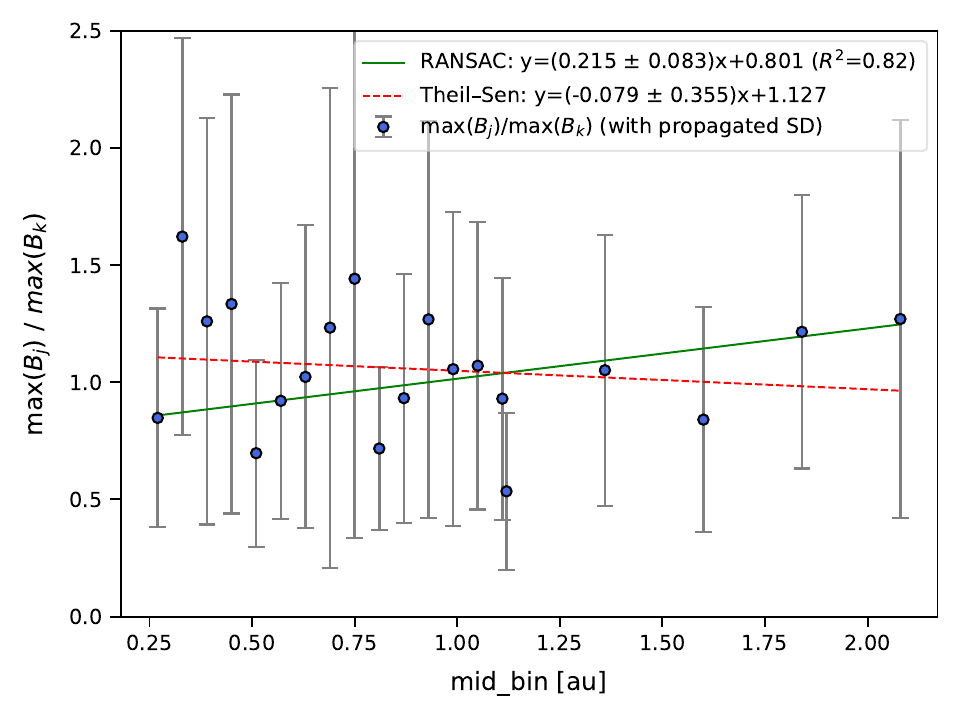}
        
    \end{minipage}
    \hfill
    \begin{minipage}[b]{0.48\linewidth}
        \centering
        \includegraphics[width=\linewidth]{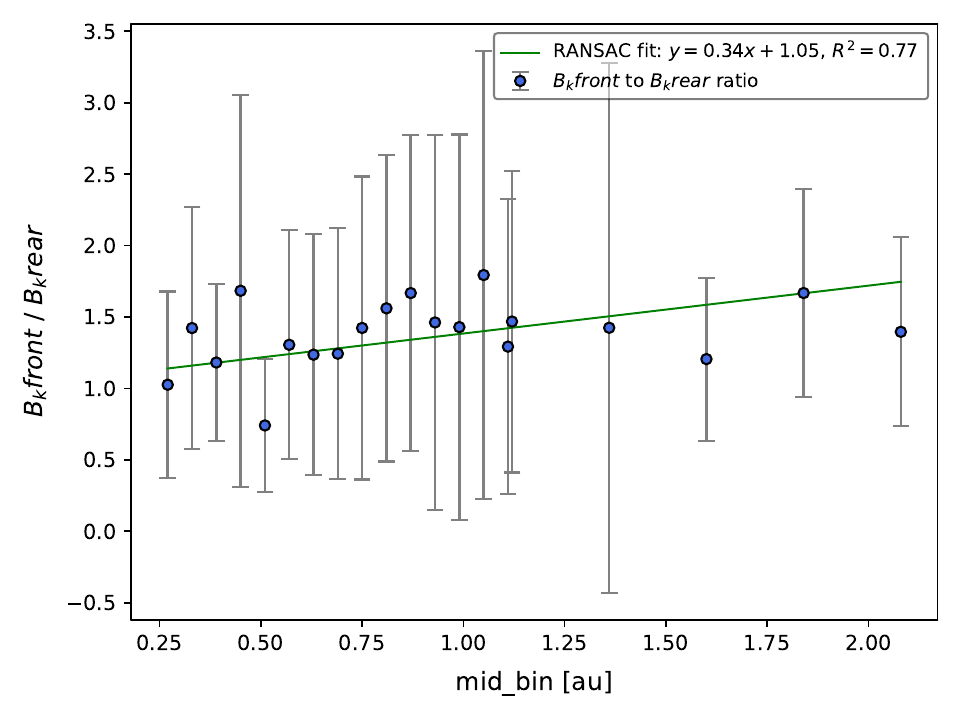} 
    \end{minipage}

    \caption{Heliocentric distance dependence of (left) the maximum toroidal-to-poloidal field ratio $\max(B_j)/\max(B_k)$ with a Theil--Sen estimator (red dotted line)  and (right) the front-to-rear ratio $B_{k,\mathrm{front}}/B_{k,\mathrm{rear}}$. In both panels, data points are plotted at the mid-bin heliocentric distance. Error bars represent the $1\sigma$ uncertainty propagated for a ratio,  $\sigma_R = R\sqrt{(\sigma_{1}/X_{1})^{2}+(\sigma_{2}/X_{2})^{2}}$, 
    where $R=X_{1}/X_{2}$ and $\sigma_{1},\sigma_{2}$ are the corresponding standard deviations of the numerator and denominator. The solid (green) lines show robust linear fits obtained with a RANSAC regressor and the fitted relations are reported in the legends as $y=mx+c$ together with the coefficient of determination $R^{2}$, computed using the RANSAC inliers only.}
    
    \label{fig:BkBj_ratio_combined}
\end{figure}

We further examine in Figure~\ref{fig:BkBj_ratio_combined}~(left) the ratio of maximum poloidal-to-toroidal magnetic field strength, $\max(B_j)/\max(B_k)$, as a function of heliocentric distance. We do so because typically the $B_k$ component has significantly different values in the front and at the rear, which may reflect aging or expansion as the CME passes over the spacecraft. Studying the ratio of the maxima, partially controls this effect. The ratio spans $\sim0.7$--$1.6$ with no apparent trend. Because the number of heliocentric distance bins is limited, linear regression can be sensitive to endpoint leverage. A RANSAC fit yields a positive slope ($0.215$~au$^{-1}$) but identifies only half of the distance bins as inliers, indicating strong sensitivity to outliers. We therefore apply the Theil--Sen estimator, which yields a slope of $-0.079$~au$^{-1}$ with a 95\% confidence interval consistent with zero. The Spearman correlation ($\rho=-0.12$, $p=0.61$) suggests no monotonic relationship between the ratio and heliocentric distance. These results indicate no statistically significant heliocentric distance dependence of the maximum poloidal-to-toroidal field ratio, implying that $B_j$ and $B_k$ decay at comparable rates during CME propagation.

The differences between the temporal and spatial profiles of CMEs have been used to highlight the CME aging effect \citep{Regnault2024a}. One approach for quantifying magnetic field asymmetry is to compute the ratio of the front-to-rear magnetic field strength, as used in \citet{Regnault2023b}. Previous studies have also examined CME aging by analyzing the expansion of magnetic field structures, such as radial expansion \citep{Nieves-Chinchilla2018, Scolini2021, Zhuang2024}. Here we adopt the front-to-rear ratio to quantify asymmetry in the internal magnetic field components as a function of heliocentric distance.

In this context, we present the front-to-rear ratio of the toroidal magnetic field across all bins as a function of heliocentric distance in Figure~\ref{fig:BkBj_ratio_combined}~(right). The ratios range from approximately 0.75 to 1.72, with most bins clustering around 1.4. This shows that for most distances the $B_k$ component of the magnetic field is not balanced but is $\sim 40\%$ stronger in the front than in the back. A linear fit to $B_k^{\rm front}/B_k^{\rm rear}$ reveals a statistically significant increasing trend with heliocentric distance (slope: $0.34 \pm 0.13$ per au, $p < 0.01$). The best-fit relation shows that the ratio is near unity at the Sun and rises to $\sim 1.4$ at 1 au. This suggests a gradual increase in internal asymmetry in this field component that may reflect the combined effects of CME expansion and aging. This differs from expectations from numerical simulations \citep[]{Regnault2024a}, which showed that aging should have a bigger impact on the measured CME properties closer to the Sun than near 1~au.

\section{Summary and Conclusion}

In this work, we investigate how the temporal properties of CMEs using more than three decades of measurements recorded by seven spacecrafts and compiled in the HELIO4CAST catalog depend on both the solar cycle and the heliocentric distance. We apply Superposed Epoch Analysis (SEA) to characterize the statistical behavior of the magnetic and plasma parameters of CMEs. To examine the solar cycle dependence, we classify the events into active phase (AP) and quiet phase (QP) groups using a sunspot number (SSN) threshold. To study the heliocentric distance dependence, we bin the events into different distance groups. This analysis is made possible because there have been over 1000 CMEs measured {\it in situ}, including over 150 measured sunward of 0.9~au, and several hundreds in quiet solar phases.

Overall, we find that physical parameters such as $B_{\mathrm{mean}}$, $T_p$, and $V_p$ of CMEs during the AP are consistently enhanced relative to the QP profiles, with the exception of the proton density ($n_p$), which remains higher in QP than in AP across the solar wind and CME substructures. To isolate the fact that CMEs in AP are typically faster than CMEs in QP from other solar cycle effects, we introduce a velocity-matching criterion that selects AP and QP events with similar propagation speeds. This step is necessary because faster CMEs typically drive stronger magnetic fields and larger kinetic energy enhancements; matching by velocity therefore removes any speed related bias. The fact that the SEA profiles for this subset closely resemble the original AP--QP profiles (just as in the non-velocity-matching  category) confirms that the observed trends are genuine intrinsic features of the CMEs rather than artifacts of speed selection.

Next, we examine the dependence of the average $B_{\mathrm{mean}}$ with heliocentric distance. We compare the resulting power law indices with those reported in previous studies, and our values show reasonable agreement, even though we considered more events and binned them by distance. Interestingly, we find significant difference in the power-law index obtained on binned or non-binned data, which might be because most CMEs have been measured close to 1~au. 

We further extend this heliocentric distance assessment to the toroidal and poloidal magnetic field components, both of which exhibit similar coefficients and exponents in their fitted parameters. This similarity is interpreted as evidence of self-similar expansion of both field components. This analysis has not been addressed in the majority of past statistical studies, so it provides a very strong observational constraint for CME flux-rope models, which should reproduce the coupled radial decay of the toroidal and poloidal field components during propagation.  

We also examine the heliocentric evolution of the maximum poloidal-to-toroidal field ratio, $\max(B_j)/\max(B_k)$, and find no statistically significant variation with heliocentric distance based on robust regression and rank-correlation tests. This result supports the interpretation that the two field components undergo comparable radial decay during propagation.

We then investigate the aging effect of CMEs resulting from their expansion during propagation by examining the front-to-rear ratio of the $B_k$ field components. These ratios show no clear trend with heliocentric distance; however, for most bins the values are approximately greater than or equal to 1. 

In this work, we present two complementary statistical analyses of multi-spacecraft CME datasets: one examining solar cycle phase dependence and the other investigating heliocentric distance dependence over 0.2--2.2~au. Together, these results highlight the diverse structural and dynamical characteristics of CME profiles across solar cycle phases and during heliocentric propagation. Further studies should look into disentangling the coupled effects of solar cycle phase and heliocentric distance on CME temporal properties as has been done for the IMF by \citet{Hanneson2020}. This will further take advantage of the thousands of {\it in situ} CME measurements to improve our understanding of CME evolution and help constrain future CME propagation models.

\section*{Data Availability statement}

The ICMECAT is available with a permanent DOI at
\url{https://doi.org/10.6084/m9.figshare.6356420}. Version 16 of the ICMECAT (1 September 2023) on figshare has been used in this study.

\begin{acknowledgments}
\section*{Acknowledgments}
We give thanks to the National Science Foundation (NSF) grant AGS1954983 and the National Aeronautics and Space Administration (NASA) under the grants 80NSSC21K0463 and 80NSSC20K0197 for supporting this research.
This work is also supported by ERC grant (HELIO4CAST, 10.3030/101042188), funded by the European Union. Views and opinions expressed are those of the author(s) only and do not necessarily reflect those of the European Union or the European Research Council Executive Agency. Neither the European Union nor the granting authority can be held responsible for them.
\end{acknowledgments}

\FloatBarrier
\clearpage
\bibliography{sample631}{}

@ARTICLE{Christian2026,
       author = {{M{\"o}stl}, Christian and {Davies}, Emma E. and {Weiler}, Eva and {Amerstorfer}, Ute V. and {Weiss}, Andreas J. and {R{\"u}disser}, Hannah T. and {Reiss}, Martin A. and {Majumdar}, Satabdwa and {Horbury}, Timothy S. and {Bale}, Stuart D. and et al.},
        title = "{On the magnetic field evolution of interplanetary coronal mass ejections from 0.07 to 5.4 au}",
      journal = {arXiv e-prints},
         year = 2026,
        month = mar,
          eid = {arXiv:2512.04730},
        pages = {arXiv:2512.04730},
          doi = {10.48550/arXiv.2512.04730},
archivePrefix = {arXiv},
       eprint = {2512.04730},
 primaryClass = {astro-ph.SR},
       adsurl = {https://ui.adsabs.harvard.edu/abs/2025arXiv251204730M}
}

@ARTICLE{Lanabere2022,
       author = {{Lanabere}, V. and {D{\'e}moulin}, P. and {Dasso}, S.},
        title = "{A robust estimation of the twist distribution in magnetic clouds}",
      journal = {\aap},
         year = 2022,
        month = dec,
       volume = {668},
          eid = {A160},
        pages = {A160},
          doi = {10.1051/0004-6361/202245062},
archivePrefix = {arXiv},
       eprint = {2211.08758},
 primaryClass = {physics.space-ph},
       adsurl = {https://ui.adsabs.harvard.edu/abs/2022A&A...668A.160L}
}

@article{Fischler1981,
  author = {Fischler, Martin A. and Bolles, Robert C},
  title     = {Random Sample Consensus: A Paradigm for Model Fitting with 
               Applications to Image Analysis and Automated Cartography},
  journal   = {Communications of the ACM},
  volume    = {24},
  number    = {6},
  pages     = {381--395},
  year      = {1981},
  month     = jun,
  doi       = {10.1145/358669.358692},
  publisher = {Association for Computing Machinery}
}

@ARTICLE{Mas-Meza2016,
       author = {{Mas{\'\i}as-Meza}, J.~J. and {Dasso}, S. and {D{\'e}moulin}, P. and {Rodriguez}, L. and {Janvier}, M.},
        title = "{Superposed epoch study of ICME sub-structures near Earth and their effects on Galactic cosmic rays}",
      journal = {\aap},
         year = 2016,
        month = aug,
       volume = {592},
          eid = {A118},
        pages = {A118},
          doi = {10.1051/0004-6361/201628571},
archivePrefix = {arXiv},
       eprint = {1605.08130},
 primaryClass = {astro-ph.SR},
       adsurl = {https://ui.adsabs.harvard.edu/abs/2016A&A...592A.118M}
}

@ARTICLE{Davies2023,
       author = {{Davies}, Emma E. and {Winslow}, R{\'e}ka M. and {Lawrence}, David J.},
        title = "{Characterizing Interplanetary Coronal Mass Ejection-related Forbush Decreases at Mercury Using MESSENGER Observations: Identification of a One- or Two-step Structure}",
      journal = {\apj},
         year = 2023,
        month = feb,
       volume = {943},
       number = {2},
          eid = {83},
        pages = {83},
          doi = {10.3847/1538-4357/acaca1},
archivePrefix = {arXiv},
       eprint = {2212.02707},
 primaryClass = {physics.space-ph},
       adsurl = {https://ui.adsabs.harvard.edu/abs/2023ApJ...943...83D}
}

@ARTICLE{Rudisser2026,
       author = {{R{\"u}disser}, Hannah T. and {Nguyen}, Gautier and {Le Lou{\"e}dec}, Justin and {Davies}, Emma E. and {M{\"o}stl}, Christian},
        title = "{ARCANE--Early Detection of Interplanetary Coronal Mass Ejections}",
      journal = {Space Weather},
         year = 2026,
        month = feb,
       volume = {24},
       number = {2},
          eid = {e2025SW004537},
        pages = {e2025SW004537},
          doi = {10.1029/2025SW004537},
archivePrefix = {arXiv},
       eprint = {2505.09365},
 primaryClass = {physics.space-ph},
       adsurl = {https://ui.adsabs.harvard.edu/abs/2026SpWea..2404537R}
}

@ARTICLE{Weiler2025,
       author = {{Weiler}, E. and {M{\"o}stl}, C. and {Davies}, E.~E. and {Veronig}, A.~M. and {Amerstorfer}, U.~V. and {Amerstorfer}, T. and {Le Lou{\"e}dec}, J. and {Bauer}, M. and {Lugaz}, N. and {Haberle}, V. and {R{\"u}disser}, H.~T. and {Majumdar}, S. and {Reiss}, M.},
        title = "{First Observations of a Geomagnetic Superstorm With a Sub-L1 Monitor}",
      journal = {Space Weather},
         year = 2025,
        month = mar,
       volume = {23},
       number = {3},
        pages = {2024SW004260},
          doi = {10.1029/2024SW004260},
archivePrefix = {arXiv},
       eprint = {2411.12490},
 primaryClass = {physics.space-ph},
       adsurl = {https://ui.adsabs.harvard.edu/abs/2025SpWea..2304260W}
}

@ARTICLE{Davies2022,
       author = {{Davies}, Emma E. and {Winslow}, R{\'e}ka M. and {Scolini}, Camilla and {Forsyth}, Robert J. and {M{\"o}stl}, Christian and {Lugaz}, No{\'e} and {Galvin}, Antoinette B.},
        title = "{Multi-spacecraft Observations of the Evolution of Interplanetary Coronal Mass Ejections between 0.3 and 2.2 au: Conjunctions with the Juno Spacecraft}",
      journal = {\apj},
         year = 2022,
        month = jul,
       volume = {933},
       number = {2},
          eid = {127},
        pages = {127},
          doi = {10.3847/1538-4357/ac731a},
archivePrefix = {arXiv},
       eprint = {2205.09472},
 primaryClass = {physics.space-ph},
       adsurl = {https://ui.adsabs.harvard.edu/abs/2022ApJ...933..127D}
}

@InCollection{Mariani1990,
  author    = {Mariani, F. and Neubauer, F. M.},
  title     = {The Interplanetary Magnetic Field},
  booktitle = {Physics of the Inner Heliosphere I},
  editor    = {Schwenn, R. and Marsch, E.},
  publisher = {Springer-Verlag},
  address   = {Berlin},
  year      = {1990},
  pages     = {183--322}
}

@ARTICLE{Osherovich1993,
       author = {{Osherovich}, V.~A. and {Farrugia}, C.~J. and {Burlaga}, L.~F.},
        title = "{Dynamics of aging magnetic clouds}",
      journal = {Advances in Space Research},
         year = 1993,
        month = jun,
       volume = {13},
       number = {6},
        pages = {57-62},
          doi = {10.1016/0273-1177(93)90391-N},
       adsurl = {https://ui.adsabs.harvard.edu/abs/1993AdSpR..13f..57O}
}

@INPROCEEDINGS{Farrugia1992,
       author = {{Farrugia}, C.~J. and {Burlaga}, L.~F. and {Osherovich}, V.~A. and {Lepping}, R.~P.},
        title = "{A comparative study of dynamically expanding force-free, constant-alpha magnetic configurations with applications to magnetic clouds}",
    booktitle = {Solar Wind Seven Colloquium},
         year = 1992,
       editor = {{Marsch}, E. and {Schwenn}, R.},
        month = jan,
        pages = {611-614},
          doi = {10.1016/B978-0-08-042049-3.50127-7},
       adsurl = {https://ui.adsabs.harvard.edu/abs/1992sws..coll..611F}
}

@ARTICLE{Farrugia1995,
       author = {{Farrugia}, C.~J. and {Erkaev}, N.~V. and {Biernat}, H.~K. and {Burlaga}, L.~F.},
        title = "{Anomalous magnetosheath properties during Earth passage of an interplanetary magnetic cloud}",
      journal = {\jgr},
         year = 1995,
        month = oct,
       volume = {100},
       number = {A10},
        pages = {19245-19258},
          doi = {10.1029/95JA01080},
       adsurl = {https://ui.adsabs.harvard.edu/abs/1995JGR...10019245F}
}

@ARTICLE{Totten1995,
       author = {{Totten}, T.~L. and {Freeman}, J.~W. and {Arya}, S.},
        title = "{An empirical determination of the polytropic index for the free-streaming solar wind using Helios 1 data}",
      journal = {\jgr},
         year = 1995,
        month = jan,
       volume = {100},
       number = {A1},
        pages = {13-18},
          doi = {10.1029/94JA02420},
       adsurl = {https://ui.adsabs.harvard.edu/abs/1995JGR...100...13T}
}

@ARTICLE{Klein1982,
       author = {{Klein}, L.~W. and {Burlaga}, L.~F.},
        title = "{Interplanetary magnetic clouds at 1 AU}",
      journal = {\jgr},
         year = 1982,
        month = feb,
       volume = {87},
       number = {A2},
        pages = {613-624},
          doi = {10.1029/JA087iA02p00613},
       adsurl = {https://ui.adsabs.harvard.edu/abs/1982JGR....87..613K}
}

@ARTICLE{Lavraud2008,
       author = {{Lavraud}, Benoit and {Borovsky}, Joseph E.},
        title = "{Altered solar wind-magnetosphere interaction at low Mach numbers: Coronal mass ejections}",
      journal = {Journal of Geophysical Research (Space Physics)},
         year = 2008,
        month = sep,
       volume = {113},
       number = {A9},
          eid = {A00B08},
        pages = {A00B08},
          doi = {10.1029/2008JA013192},
       adsurl = {https://ui.adsabs.harvard.edu/abs/2008JGRA..113.0B08L}
}

@ARTICLE{Farrugia1993,
       author = {{Farrugia}, C.~J. and {Burlaga}, L.~F. and {Osherovich}, V.~A. and {Richardson}, I.~G. and {Freeman}, M.~P. and {Lepping}, R.~P. and {Lazarus}, A.~J.},
        title = "{A study of an expanding interplanetary magnetic cloud and its interaction with the Earth's magnetosphere: The interplanetary aspect}",
      journal = {\jgr},
         year = 1993,
        month = may,
       volume = {98},
       number = {A5},
        pages = {7621-7632},
          doi = {10.1029/92JA02349},
       adsurl = {https://ui.adsabs.harvard.edu/abs/1993JGR....98.7621F}
}

@ARTICLE{Gulisano2010,
       author = {{Gulisano}, A.~M. and {D{\'e}moulin}, P. and {Dasso}, S. and {Ruiz}, M.~E. and {Marsch}, E.},
        title = "{Global and local expansion of magnetic clouds in the inner heliosphere}",
      journal = {\aap},
         year = 2010,
        month = jan,
       volume = {509},
          eid = {A39},
        pages = {A39},
          doi = {10.1051/0004-6361/200912375},
archivePrefix = {arXiv},
       eprint = {1206.1112},
 primaryClass = {physics.space-ph},
       adsurl = {https://ui.adsabs.harvard.edu/abs/2010A&A...509A..39G}
}

@ARTICLE{Nieves-Chinchilla2018,
       author = {{Nieves-Chinchilla}, T. and {Vourlidas}, A. and {Raymond}, J.~C. and {Linton}, M.~G. and {Al-haddad}, N. and {Savani}, N.~P. and {Szabo}, A. and {Hidalgo}, M.~A.},
        title = "{Understanding the Internal Magnetic Field Configurations of ICMEs Using More than 20 Years of Wind Observations}",
      journal = {\solphys},
         year = 2018,
        month = feb,
       volume = {293},
       number = {2},
          eid = {25},
        pages = {25},
          doi = {10.1007/s11207-018-1247-z},
       adsurl = {https://ui.adsabs.harvard.edu/abs/2018SoPh..293...25N}
}

@ARTICLE{Hanneson2020,
       author = {{Hanneson}, Cedar and {Johnson}, Catherine L. and {Mittelholz}, Anna and {Al Asad}, Manar M. and {Goldblatt}, Colin},
        title = "{Dependence of the Interplanetary Magnetic Field on Heliocentric Distance at 0.3-1.7 AU: A Six-Spacecraft Study}",
      journal = {Journal of Geophysical Research (Space Physics)},
         year = 2020,
        month = mar,
       volume = {125},
       number = {3},
          eid = {e27139},
        pages = {e27139},
          doi = {10.1029/2019JA027139},
       adsurl = {https://ui.adsabs.harvard.edu/abs/2020JGRA..12527139H}
}

@ARTICLE{Moestl2014,
       author = {{M{\"o}stl}, C. and {Amla}, K. and {Hall}, J.~R. and {Liewer}, P.~C. and {De Jong}, E.~M. and {Colaninno}, R.~C. and {Veronig}, A.~M. and {Rollett}, T. and {Temmer}, M. and {Peinhart}, V. and {Davies}, J.~A. and {Lugaz}, N. and {Liu}, Y.~D. and {Farrugia}, C.~J. and {Luhmann}, J.~G. and {Vr{\v{s}}nak}, B. and {Harrison}, R.~A. and {Galvin}, A.~B.},
        title = "{Connecting Speeds, Directions and Arrival Times of 22 Coronal Mass Ejections from the Sun to 1 AU}",
      journal = {\apj},
         year = 2014,
        month = jun,
       volume = {787},
       number = {2},
          eid = {119},
        pages = {119},
          doi = {10.1088/0004-637X/787/2/119},
archivePrefix = {arXiv},
       eprint = {1404.3579},
 primaryClass = {astro-ph.SR},
       adsurl = {https://ui.adsabs.harvard.edu/abs/2014ApJ...787..119M}
}

@INPROCEEDINGS{Tousey1973,
       author = {{Tousey}, R.},
        title = "{The solar corona.}",
    booktitle = {Space Research Conference},
         year = 1973,
       editor = {{Rycroft}, M.~J. and {Runcorn}, S.~K.},
       volume = {2},
        month = jan,
        pages = {713-730},
       adsurl = {https://ui.adsabs.harvard.edu/abs/1973spre.conf..713T}
}

@ARTICLE{Muller2020,
       author = {{M{\"u}ller}, D. and {St. Cyr}, O.~C. and {Zouganelis}, I. and {Gilbert}, H.~R. and {Marsden}, R. and {Nieves-Chinchilla}, T. and {Antonucci}, E. and {Auch{\`e}re}, F. and {Berghmans}, D. and {Horbury}, T.~S. and {Howard}, R.~A. and {Krucker}, S. and {Maksimovic}, M. and {Owen}, C.~J. and {Rochus}, P. and {Rodriguez-Pacheco}, J. and {Romoli}, M. and {Solanki}, S.~K. and {Bruno}, R. and {Carlsson}, M. and {Fludra}, A. and {Harra}, L. and {Hassler}, D.~M. and {Livi}, S. and {Louarn}, P. and {Peter}, H. and {Sch{\"u}hle}, U. and {Teriaca}, L. and {del Toro Iniesta}, J.~C. and {Wimmer-Schweingruber}, R.~F. and {Marsch}, E. and {Velli}, M. and {De Groof}, A. and {Walsh}, A. and {Williams}, D.},
        title = "{The Solar Orbiter mission. Science overview}",
      journal = {\aap},
         year = 2020,
        month = oct,
       volume = {642},
          eid = {A1},
        pages = {A1},
          doi = {10.1051/0004-6361/202038467},
archivePrefix = {arXiv},
       eprint = {2009.00861},
 primaryClass = {astro-ph.SR},
       adsurl = {https://ui.adsabs.harvard.edu/abs/2020A&A...642A...1M}
}

@ARTICLE{Regnault2024a,
       author = {{Regnault}, F. and {Al-Haddad}, N. and {Lugaz}, N. and {Farrugia}, C.~J. and {Zhuang}, B. and {Yu}, W. and {Strugarek}, A.},
        title = "{Exploring the Impact of the Aging Effect on Inferred Properties of Solar Coronal Mass Ejections}",
      journal = {\apjl},
         year = 2024,
        month = may,
       volume = {966},
       number = {1},
          eid = {L17},
        pages = {L17},
          doi = {10.3847/2041-8213/ad3806},
       adsurl = {https://ui.adsabs.harvard.edu/abs/2024ApJ...966L..17R}
}

@ARTICLE{Zhuang2024,
       author = {{Zhuang}, B. and {Lugaz}, N. and {Al-Haddad}, N. and {Scolini}, C. and {Farrugia}, C.~J. and {Regnault}, F. and {Davies}, E.~E. and {Yu}, W. and {Winslow}, R.~M. and {Galvin}, A.~B.},
        title = "{Combining STEREO heliospheric imagers and Solar Orbiter to investigate the evolution of the 2022 March 10 CME}",
      journal = {\aap},
         year = 2024,
        month = feb,
       volume = {682},
          eid = {A107},
        pages = {A107},
          doi = {10.1051/0004-6361/202347561},
       adsurl = {https://ui.adsabs.harvard.edu/abs/2024A&A...682A.107Z}
}

@ARTICLE{Scolini2021,
       author = {{Scolini}, C. and {Dasso}, S. and {Rodriguez}, L. and {Zhukov}, A.~N. and {Poedts}, S.},
        title = "{Exploring the radial evolution of interplanetary coronal mass ejections using EUHFORIA}",
      journal = {\aap},
         year = 2021,
        month = may,
       volume = {649},
          eid = {A69},
        pages = {A69},
          doi = {10.1051/0004-6361/202040226},
archivePrefix = {arXiv},
       eprint = {2102.07569},
 primaryClass = {astro-ph.SR},
       adsurl = {https://ui.adsabs.harvard.edu/abs/2021A&A...649A..69S}
}

@ARTICLE{Regnault2023b,
       author = {{Regnault}, F. and {Al-Haddad}, N. and {Lugaz}, N. and {Farrugia}, C.~J. and {Yu}, W. and {Davies}, E.~E. and {Galvin}, A.~B. and {Zhuang}, B.},
        title = "{Investigating the Magnetic Structure of Interplanetary Coronal Mass Ejections Using Simultaneous Multispacecraft In Situ Measurements}",
      journal = {\apj},
         year = 2023,
        month = nov,
       volume = {957},
       number = {1},
          eid = {49},
        pages = {49},
          doi = {10.3847/1538-4357/acef16},
archivePrefix = {arXiv},
       eprint = {2309.10582},
 primaryClass = {astro-ph.SR},
       adsurl = {https://ui.adsabs.harvard.edu/abs/2023ApJ...957...49R}
}

@ARTICLE{Berdichevsky2003,
       author = {{Berdichevsky}, D.~B. and {Lepping}, R.~P. and {Farrugia}, C.~J.},
        title = "{Geometric considerations of the evolution of magnetic flux ropes}",
      journal = {\pre},
         year = 2003,
        month = mar,
       volume = {67},
       number = {3},
          eid = {036405},
        pages = {036405},
          doi = {10.1103/PhysRevE.67.036405},
       adsurl = {https://ui.adsabs.harvard.edu/abs/2003PhRvE..67c6405B}
}

@ARTICLE{Shimazu2002,
       author = {{Shimazu}, Hironori and {Vandas}, Marek},
        title = "{A self-similar solution of expanding cylindrical flux ropes for any polytropic index value}",
      journal = {Earth, Planets and Space},
         year = 2002,
        month = jul,
       volume = {54},
       number = {7},
        pages = {783-790},
          doi = {10.1186/BF03351731},
       adsurl = {https://ui.adsabs.harvard.edu/abs/2002EP&S...54..783S}
}

@ARTICLE{Leitner2007,
       author = {{Leitner}, M. and {Farrugia}, C.~J. and {M{\"o}stl}, C. and {Ogilvie}, K.~W. and {Galvin}, A.~B. and {Schwenn}, R. and {Biernat}, H.~K.},
        title = "{Consequences of the force-free model of magnetic clouds for their heliospheric evolution}",
      journal = {Journal of Geophysical Research (Space Physics)},
         year = 2007,
        month = jun,
       volume = {112},
       number = {A6},
          eid = {A06113},
        pages = {A06113},
          doi = {10.1029/2006JA011940},
       adsurl = {https://ui.adsabs.harvard.edu/abs/2007JGRA..112.6113L}
}

@ARTICLE{Siscoe2008,
       author = {{Siscoe}, G. and {Odstrcil}, D.},
        title = "{Ways in which ICME sheaths differ from magnetosheaths}",
      journal = {Journal of Geophysical Research (Space Physics)},
         year = 2008,
        month = sep,
       volume = {113},
       number = {A9},
          eid = {A00B07},
        pages = {A00B07},
          doi = {10.1029/2008JA013142},
       adsurl = {https://ui.adsabs.harvard.edu/abs/2008JGRA..113.0B07S}
}

@ARTICLE{weiss_2024,
       author = {{Weiss}, Andreas J. and {Nieves-Chinchilla}, Teresa and {M{\"o}stl}, Christian},
        title = "{Distorted Magnetic Flux Ropes within Interplanetary Coronal Mass Ejections}",
      journal = {\apj},
         year = 2024,
        month = nov,
       volume = {975},
       number = {2},
          eid = {169},
        pages = {169},
          doi = {10.3847/1538-4357/ad7940},
archivePrefix = {arXiv},
       eprint = {2406.13022},
 primaryClass = {astro-ph.SR},
       adsurl = {https://ui.adsabs.harvard.edu/abs/2024ApJ...975..169W}
}

@Article{Winslow2015,
  author = {Winslow, Reka M. and Lugaz, No{\'e} and Philpott, Lydia C. and Schwadron, Nathan A. and Farrugia, Charles J. and Anderson, Brian J. and Smith, Charles W.},
  title = {Interplanetary coronal mass ejections from MESSENGER orbital observations at Mercury},
  journal = {Journal of Geophysical Research: Space Physics},
  volume = {120},
  pages = {6101--6118},
  year = {2015}
}

@Article{Gopalswamy2016,
  author  = {Gopalswamy, N.},
  title   = {History and Development of Coronal Mass Ejections as a Key Player in Solar Terrestrial Relationship},
  journal = {Geoscience Letters},
  pages   = {1--36},
  year    = {2016}
}

@Article{Gosling1993,
  author  = {Gosling, J. T.},
  title   = {The solar flare myth},
  journal = {Journal of Geophysical Research},
  volume  = {98},
  pages   = {18937--18949},
  year    = {1993}
}

@Article{Wilson1987,
  author  = {Wilson, R. M.},
  title   = {Geomagnetic response to magnetic clouds},
  journal = {Planetary and Space Science},
  volume  = {35},
  year    = {1987}
}

@Article{Christian2017,
  author  = {Moestl, C. and Isavnin, A. and Boakes, P. D. and Kilpua, E. K. J. and Davies, J. A. and Harrison, R. A. and Barnes, D. and Krupar, V. and Eastwood, J. P. and Good, S. W. and Forsyth, R. J. and Bothmer, V. and Reiss, M. A. and Amerstorfer, T. and Winslow, R. M. and Anderson, B. J. and Philpott, L. C. and Rodriguez, L. and Rouillard, A. P. and Gallagher, P. and Nieves-Chinchilla, T. and Zhang, T. L.},
  title   = {Modeling observations of solar coronal mass ejections with heliospheric imagers verified with the Heliophysics System Observatory},
  journal = {Space Weather},
  volume  = {15},
  pages   = {955--970},
  year    = {2017},
  doi     = {10.1002/2017SW001614}
}

@Article{Christian2020,
  author = {Moestl, Christian and Weiss, Andreas J. and Bailey, Rachel L. and Reiss, Martin A. and Amerstorfer, Tanja and Hinterreiter, Juergen and Bauer, Maike and McIntosh, Scott W. and Lugaz, No{\'e} and Stansby, David},
  title = {Prediction of the In Situ Coronal Mass Ejection Rate for Solar Cycle 25: Implications for Parker Solar Probe In Situ Observations},
  journal = {The Astrophysical Journal},
  volume = {44},
  pages = {1--9},
  year = {2020}
}

@ARTICLE{Connerney2017,
       author = {{Connerney}, J.~E.~P. and {Benn}, M. and {Bjarno}, J.~B. and {Denver}, T. and {Espley}, J. and {Jorgensen}, J.~L. and {Jorgensen}, P.~S. and {Lawton}, P. and {Malinnikova}, A. and {Merayo}, J.~M. and {Murphy}, S. and {Odom}, J. and {Oliversen}, R. and {Schnurr}, R. and {Sheppard}, D. and {Smith}, E.~J.},
        title = "{The Juno Magnetic Field Investigation}",
      journal = {\ssr},
         year = 2017,
        month = nov,
       volume = {213},
       number = {1-4},
        pages = {39-138},
          doi = {10.1007/s11214-017-0334-z},
       adsurl = {https://ui.adsabs.harvard.edu/abs/2017SSRv..213...39C}
}

@ARTICLE{Janvier2019,
       author = {{Janvier}, Miho and {Winslow}, Reka M. and {Good}, Simon and {Bonhomme}, Elise and {D{\'e}moulin}, Pascal and {Dasso}, Sergio and {M{\"o}stl}, Christian and {Lugaz}, No{\'e} and {Amerstorfer}, Tanja and {Soubri{\'e}}, Elie and {Boakes}, Peter D.},
        title = "{Generic Magnetic Field Intensity Profiles of Interplanetary Coronal Mass Ejections at Mercury, Venus, and Earth From Superposed Epoch Analyses}",
      journal = {Journal of Geophysical Research (Space Physics)},
         year = 2019,
        month = feb,
       volume = {124},
       number = {2},
        pages = {812-836},
          doi = {10.1029/2018JA025949},
archivePrefix = {arXiv},
       eprint = {1901.09921},
 primaryClass = {physics.space-ph},
       adsurl = {https://ui.adsabs.harvard.edu/abs/2019JGRA..124..812J}
}

@INPROCEEDINGS{Regnault2020,
       author = {{Regnault}, Florian and {Dasso}, Sergio and {Auchere}, Frederic and {Demoulin}, Pascal and {Janvier}, Miho and {Strugarek}, Antoine},
        title = "{20 years of ACE data: how superposed epoch analyses reveal generic features in interplanetary CME profiles}",
    booktitle = {43rd COSPAR Scientific Assembly. Held 28 January - 4 February},
         year = 2021,
       volume = {43},
        month = jan,
          eid = {1017},
        pages = {1017},
       adsurl = {https://ui.adsabs.harvard.edu/abs/2021cosp...43E1017R}
}

@ARTICLE{Yermolaev2020,
       author = {{Yermolaev}, Y.~I. and {Lodkina}, I.~G. and {Yermolaev}, M.~Y. and {Riazantseva}, M.~O. and {Rakhmanova}, L.~S. and {Borodkova}, N.~L. and {Shugay}, Y.~S. and {Slemzin}, V.~A. and {Veselovsky}, I.~S. and {Rodkin}, D.~G.},
        title = "{Dynamics of Large-Scale Solar-Wind Streams Obtained by the Double Superposed Epoch Analysis: 4. Helium Abundance}",
      journal = {Journal of Geophysical Research (Space Physics)},
         year = 2020,
        month = jul,
       volume = {125},
       number = {7},
          eid = {e27878},
        pages = {e27878},
          doi = {10.1029/2020JA027878},
archivePrefix = {arXiv},
       eprint = {1807.03579},
 primaryClass = {physics.space-ph}
}

@ARTICLE{Rodriguez2016,
       author = {{Rodriguez}, L. and {Mas{\'\i}as-Meza}, J.~J. and {Dasso}, S. and {D{\'e}moulin}, P. and {Zhukov}, A.~N. and {Gulisano}, A.~M. and {Mierla}, M. and {Kilpua}, E. and {West}, M. and {Lacatus}, D. and {Paraschiv}, A. and {Janvier}, M.},
        title = "{Typical Profiles and Distributions of Plasma and Magnetic Field Parameters in Magnetic Clouds at 1 AU}",
      journal = {\solphys},
         year = 2016,
        month = aug,
       volume = {291},
       number = {7},
        pages = {2145-2163},
          doi = {10.1007/s11207-016-0955-5},
       adsurl = {https://ui.adsabs.harvard.edu/abs/2016SoPh..291.2145R}
}

@ARTICLE{Davies2021,
       author = {{Davies}, Emma E. and {Forsyth}, Robert J. and {Winslow}, R{\'e}ka M. and {M{\"o}stl}, Christian and {Lugaz}, No{\'e}},
        title = "{A Catalog of Interplanetary Coronal Mass Ejections Observed by Juno between 1 and 5.4 au}",
      journal = {\apj},
         year = 2021,
        month = dec,
       volume = {923},
       number = {2},
          eid = {136},
        pages = {136},
          doi = {10.3847/1538-4357/ac2ccb},
archivePrefix = {arXiv},
       eprint = {2111.11336},
 primaryClass = {physics.space-ph},
       adsurl = {https://ui.adsabs.harvard.edu/abs/2021ApJ...923..136D}
}

@ARTICLE{Solomon2001,
       author = {{Solomon}, Sean C. and {McNutt}, Ralph L. and {Gold}, Robert E. and {Acu{\~n}a}, Mario H. and {Baker}, Daniel N. and {Boynton}, William V. and {Chapman}, Clark R. and {Cheng}, Andrew F. and {Gloeckler}, George and {Head}, III, James W. and {Krimigis}, Stamatios M. and {McClintock}, William E. and {Murchie}, Scott L. and {Peale}, Stanton J. and {Phillips}, Roger J. and {Robinson}, Mark S. and {Slavin}, James A. and {Smith}, David E. and {Strom}, Robert G. and {Trombka}, Jacob I. and {Zuber}, Maria T.},
        title = "{The MESSENGER mission to Mercury: scientific objectives and implementation}",
      journal = {\planss},
         year = 2001,
        month = dec,
       volume = {49},
       number = {14-15},
        pages = {1445-1465},
          doi = {10.1016/S0032-0633(01)00085-X},
       adsurl = {https://ui.adsabs.harvard.edu/abs/2001P&SS...49.1445S}
}

@ARTICLE{Stone1998,
       author = {{Stone}, E.~C. and {Frandsen}, A.~M. and {Mewaldt}, R.~A. and {Christian}, E.~R. and {Margolies}, D. and {Ormes}, J.~F. and {Snow}, F.},
        title = "{The Advanced Composition Explorer}",
      journal = {\ssr},
         year = 1998,
        month = jul,
       volume = {86},
        pages = {1-22},
          doi = {10.1023/A:1005082526237},
       adsurl = {https://ui.adsabs.harvard.edu/abs/1998SSRv...86....1S}
}

@ARTICLE{Salman2020,
       author = {{Salman}, Tarik M. and {Lugaz}, No{\'e} and {Farrugia}, Charles J. and {Winslow}, Reka M. and {Jian}, Lan K. and {Galvin}, Antoinette B.},
        title = "{Properties of the Sheath Regions of Coronal Mass Ejections with or without Shocks from STEREO in situ Observations near 1 au}",
      journal = {\apj},
         year = 2020,
        month = dec,
       volume = {904},
       number = {2},
          eid = {177},
        pages = {177},
          doi = {10.3847/1538-4357/abbdf5},
archivePrefix = {arXiv},
       eprint = {2011.06632},
 primaryClass = {physics.space-ph},
       adsurl = {https://ui.adsabs.harvard.edu/abs/2020ApJ...904..177S}
}

@ARTICLE{Richardson1995,
       author = {{Richardson}, I.~G. and {Cane}, H.~V.},
        title = "{Regions of abnormally low proton temperature in the solar wind (1965-1991) and their association with ejecta}",
      journal = {\jgr},
         year = 1995,
        month = dec,
       volume = {100},
       number = {A12},
        pages = {23397-23412},
          doi = {10.1029/95JA02684},
       adsurl = {https://ui.adsabs.harvard.edu/abs/1995JGR...10023397R}
}

@ARTICLE{Gosling1986,
       author = {{Gosling}, J.~T. and {Baker}, D.~N. and {Bame}, S.~J. and {Zwickl}, R.~D.},
        title = "{Bidirectional solar wind electron heat flux and hemispherically symmetric polar rain}",
      journal = {\jgr},
         year = 1986,
        month = oct,
       volume = {91},
       number = {A10},
        pages = {11352-11358},
          doi = {10.1029/JA091iA10p11352},
       adsurl = {https://ui.adsabs.harvard.edu/abs/1986JGR....9111352G}
}

@ARTICLE{Gosling1973,
       author = {{Gosling}, J.~T. and {Pizzo}, V. and {Bame}, S.~J.},
        title = "{Anomalously low proton temperatures in the solar wind following interplanetary shock waves{\textemdash}evidence for magnetic bottles?}",
      journal = {\jgr},
         year = 1973,
        month = jan,
       volume = {78},
       number = {13},
        pages = {2001},
          doi = {10.1029/JA078i013p02001},
       adsurl = {https://ui.adsabs.harvard.edu/abs/1973JGR....78.2001G}
}

@ARTICLE{Wang2005,
       author = {{Wang}, C. and {Du}, D. and {Richardson}, J.~D.},
        title = "{Characteristics of the interplanetary coronal mass ejections in the heliosphere between 0.3 and 5.4 AU}",
      journal = {Journal of Geophysical Research (Space Physics)},
         year = 2005,
        month = oct,
       volume = {110},
       number = {A10},
          eid = {A10107},
        pages = {A10107},
          doi = {10.1029/2005JA011198},
       adsurl = {https://ui.adsabs.harvard.edu/abs/2005JGRA..11010107W}
}

@ARTICLE{Wimmer-Schweingruber2006,
       author = {{Wimmer-Schweingruber}, R.~F. and {Crooker}, N.~U. and {Balogh}, A. and {Bothmer}, V. and {Forsyth}, R.~J. and {Gazis}, P. and {Gosling}, J.~T. and {Horbury}, T. and {Kilchenmann}, A. and {Richardson}, I.~G. and {Richardson}, J.~D. and {Riley}, P. and {Rodriguez}, L. and {von Steiger}, R. and {Wurz}, P. and {Zurbuchen}, T.~H.},
        title = "{Understanding Interplanetary Coronal Mass Ejection Signatures.  Report of Working Group B}",
      journal = {\ssr},
         year = 2006,
        month = mar,
       volume = {123},
       number = {1-3},
        pages = {177-216},
          doi = {10.1007/s11214-006-9017-x},
       adsurl = {https://ui.adsabs.harvard.edu/abs/2006SSRv..123..177W}
}

@ARTICLE{Zurbuchen2006,
       author = {{Zurbuchen}, Thomas H. and {Richardson}, Ian G.},
        title = "{In-Situ Solar Wind and Magnetic Field Signatures of Interplanetary Coronal Mass Ejections}",
      journal = {\ssr},
         year = 2006,
        month = mar,
       volume = {123},
       number = {1-3},
        pages = {31-43},
          doi = {10.1007/s11214-006-9010-4},
       adsurl = {https://ui.adsabs.harvard.edu/abs/2006SSRv..123...31Z}
}

@ARTICLE{Borrini1982,
       author = {{Borrini}, G. and {Gosling}, J.~T. and {Bame}, S.~J. and {Feldman}, W.~C.},
        title = "{Helium abundance enhancements in the solar wind}",
      journal = {\jgr},
         year = 1982,
        month = sep,
       volume = {87},
       number = {A9},
        pages = {7370-7378},
          doi = {10.1029/JA087iA09p07370},
       adsurl = {https://ui.adsabs.harvard.edu/abs/1982JGR....87.7370B}
}

@ARTICLE{Burlaga1981,
       author = {{Burlaga}, L. and {Sittler}, E. and {Mariani}, F. and {Schwenn}, R.},
        title = "{Magnetic loop behind an interplanetary shock: Voyager, Helios, and IMP 8 observations}",
      journal = {\jgr},
         year = 1981,
        month = aug,
       volume = {86},
       number = {A8},
        pages = {6673-6684},
          doi = {10.1029/JA086iA08p06673},
       adsurl = {https://ui.adsabs.harvard.edu/abs/1981JGR....86.6673B}
}

@ARTICLE{Lugaz2020,
       author = {{Lugaz}, No{\'e} and {Salman}, Tarik M. and {Winslow}, R{\'e}ka M. and {Al-Haddad}, Nada and {Farrugia}, Charles J. and {Zhuang}, Bin and {Galvin}, Antoinette B.},
        title = "{Inconsistencies Between Local and Global Measures of CME Radial Expansion as Revealed by Spacecraft Conjunctions}",
      journal = {\apj},
         year = 2020,
        month = aug,
       volume = {899},
       number = {2},
          eid = {119},
        pages = {119},
          doi = {10.3847/1538-4357/aba26b},
archivePrefix = {arXiv},
       eprint = {2007.01699},
 primaryClass = {physics.space-ph},
       adsurl = {https://ui.adsabs.harvard.edu/abs/2020ApJ...899..119L}
}

@ARTICLE{Banu2024,
       author = {{Banu}, Sahanaj A. and {Winslow}, R{\'e}ka M. and {Scolini}, Camilla and {Davies}, Emma E. and {Farrugia}, Charles J. and {Murphy}, Amy K. and {Lugaz}, No{\'e} and {Al-Haddad}, Nada},
        title = "{Characterization of Small Flux Ropes Using Juno Spacecraft Cruise-phase Data}",
      journal = {\apj},
         year = 2024,
        month = nov,
       volume = {976},
       number = {1},
          eid = {79},
        pages = {79},
          doi = {10.3847/1538-4357/ad833e},
       adsurl = {https://ui.adsabs.harvard.edu/abs/2024ApJ...976...79B}
}

@ARTICLE{Yu2024,
       author = {{Yu}, Wenyuan and {Al-Haddad}, Nada and {Farrugia}, Charles J. and {Lugaz}, No{\'e} and {Zhuang}, Bin and {Regnault}, Florian and {Galvin}, Antoinette B.},
        title = "{Measurements of Magnetic Cloud Expansion through Multiple Spacecraft in Radial Conjunction}",
      journal = {\apj},
         year = 2024,
        month = oct,
       volume = {974},
       number = {2},
          eid = {289},
        pages = {289},
          doi = {10.3847/1538-4357/ad71cc},
       adsurl = {https://ui.adsabs.harvard.edu/abs/2024ApJ...974..289Y}
}

@ARTICLE{Cartwright2010,
       author = {{Cartwright}, M.~L. and {Moldwin}, M.~B.},
        title = "{Heliospheric evolution of solar wind small-scale magnetic flux ropes}",
      journal = {Journal of Geophysical Research (Space Physics)},
         year = 2010,
        month = aug,
       volume = {115},
       number = {A8},
          eid = {A08102},
        pages = {A08102},
          doi = {10.1029/2009JA014271},
       adsurl = {https://ui.adsabs.harvard.edu/abs/2010JGRA..115.8102C}
}

@ARTICLE{Winslow2013,
       author = {{Winslow}, Reka M. and {Anderson}, Brian J. and {Johnson}, Catherine L. and {Slavin}, James A. and {Korth}, Haje and {Purucker}, Michael E. and {Baker}, Daniel N. and {Solomon}, Sean C.},
        title = "{Mercury's magnetopause and bow shock from MESSENGER Magnetometer observations}",
      journal = {Journal of Geophysical Research (Space Physics)},
         year = 2013,
        month = may,
       volume = {118},
       number = {5},
        pages = {2213-2227},
          doi = {10.1002/jgra.50237},
       adsurl = {https://ui.adsabs.harvard.edu/abs/2013JGRA..118.2213W}
}

@ARTICLE{Salman2020a,
       author = {{Salman}, T.~M. and {Winslow}, R.~M. and {Lugaz}, N.},
        title = "{Radial Evolution of Coronal Mass Ejections Between MESSENGER, Venus Express, STEREO, and L1: Catalog and Analysis}",
      journal = {Journal of Geophysical Research (Space Physics)},
         year = 2020,
        month = jan,
       volume = {125},
       number = {1},
          eid = {e27084},
        pages = {e27084},
          doi = {10.1029/2019JA027084},
archivePrefix = {arXiv},
       eprint = {1912.11731},
 primaryClass = {physics.space-ph},
       adsurl = {https://ui.adsabs.harvard.edu/abs/2020JGRA..12527084S}
}

@ARTICLE{Song2021,
       author = {{Song}, Hongqiang and {Li}, Leping and {Sun}, Yanyan and {Lv}, Qi and {Zheng}, Ruisheng and {Chen}, Yao},
        title = "{Solar Cycle Dependence of ICME Composition}",
      journal = {\solphys},
         year = 2021,
        month = jul,
       volume = {296},
       number = {7},
          eid = {111},
        pages = {111},
          doi = {10.1007/s11207-021-01852-y},
archivePrefix = {arXiv},
       eprint = {2106.03003},
 primaryClass = {astro-ph.SR},
       adsurl = {https://ui.adsabs.harvard.edu/abs/2021SoPh..296..111S}
}

@ARTICLE{Good2016,
       author = {{Good}, S.~W. and {Forsyth}, R.~J.},
        title = "{Interplanetary Coronal Mass Ejections Observed by MESSENGER and Venus Express}",
      journal = {\solphys},
         year = 2016,
        month = jan,
       volume = {291},
       number = {1},
        pages = {239-263},
          doi = {10.1007/s11207-015-0828-3},
archivePrefix = {arXiv},
       eprint = {1511.07749},
 primaryClass = {physics.space-ph},
       adsurl = {https://ui.adsabs.harvard.edu/abs/2016SoPh..291..239G}
}

@ARTICLE{Regnault2024,
       author = {{Regnault}, F. and {Al-Haddad}, N. and {Lugaz}, N. and {Farrugia}, C.~J. and {Yu}, W. and {Zhuang}, B. and {Davies}, E.~E.},
        title = "{Discrepancies in the Properties of a Coronal Mass Ejection on Scales of 0.03 au as Revealed by Simultaneous Measurements at Solar Orbiter and Wind: The 2021 November 3{\textendash}5 Event}",
      journal = {\apj},
         year = 2024,
        month = feb,
       volume = {962},
       number = {2},
          eid = {190},
        pages = {190},
          doi = {10.3847/1538-4357/ad1883},
archivePrefix = {arXiv},
       eprint = {2311.14046},
 primaryClass = {astro-ph.SR},
       adsurl = {https://ui.adsabs.harvard.edu/abs/2024ApJ...962..190R}
}

@ARTICLE{Vourlidas2010,
       author = {{Vourlidas}, A. and {Howard}, R.~A. and {Esfandiari}, E. and {Patsourakos}, S. and {Yashiro}, S. and {Michalek}, G.},
        title = "{Comprehensive Analysis of Coronal Mass Ejection Mass and Energy Properties Over a Full Solar Cycle}",
      journal = {\apj},
         year = 2010,
        month = oct,
       volume = {722},
       number = {2},
        pages = {1522-1538},
          doi = {10.1088/0004-637X/722/2/1522},
archivePrefix = {arXiv},
       eprint = {1008.3737},
 primaryClass = {astro-ph.SR},
       adsurl = {https://ui.adsabs.harvard.edu/abs/2010ApJ...722.1522V}
}

@INPROCEEDINGS{Hundhausen1999,
       author = {{Hundhausen}, A.},
        title = "{Coronal Mass Ejections}",
    booktitle = {The many faces of the sun: a summary of the results from NASA's Solar Maximum Mission.},
         year = 1999,
       editor = {{Strong}, Keith T. and {Saba}, Julia L.~R. and {Haisch}, Bernhard M. and {Schmelz}, Joan T.},
        month = jan,
        pages = {143},
       adsurl = {https://ui.adsabs.harvard.edu/abs/1999mfs..conf..143H}
}

@ARTICLE{Hutchinson2011,
       author = {{Hutchinson}, J.~A. and {Wright}, D.~M. and {Milan}, S.~E.},
        title = "{Geomagnetic storms over the last solar cycle: A superposed epoch analysis}",
      journal = {Journal of Geophysical Research (Space Physics)},
         year = 2011,
        month = sep,
       volume = {116},
       number = {A9},
          eid = {A09211},
        pages = {A09211},
          doi = {10.1029/2011JA016463},
       adsurl = {https://ui.adsabs.harvard.edu/abs/2011JGRA..116.9211H}
}

@ARTICLE{Cremades2007,
       author = {{Cremades}, Hebe and {St. Cyr}, O.~C.},
        title = "{Coronal mass ejections: Solar cycle aspects}",
      journal = {Advances in Space Research},
         year = 2007,
        month = jan,
       volume = {40},
       number = {7},
        pages = {1042-1048},
          doi = {10.1016/j.asr.2007.01.088},
       adsurl = {https://ui.adsabs.harvard.edu/abs/2007AdSpR..40.1042C}
}

@InCollection{Alexander2006,
  author    = {Alexander, David and Richardson, Ian G. and Zurbuchen, Thomas H.},
  title     = {A Brief History of CME Science},
  booktitle = {Coronal Mass Ejections},
  editor    = {Kunow, H. and Crooker, N. U. and Linker, J. A. and Schwenn, R. and von Steiger, R.},
  publisher = {Springer},
  volume    = {21},
  pages     = {3--21},
  year      = {2006},
  doi       = {10.1007/978-0-387-45088-9_1}
}

@ARTICLE{Zhang2006,
       author = {{Zhang}, T.~L. and {Baumjohann}, W. and {Delva}, M. and {Auster}, H. -U. and {Balogh}, A. and {Russell}, C.~T. and {Barabash}, S. and {Balikhin}, M. and {Berghofer}, G. and {Biernat}, H.~K. and {Lammer}, H. and {Lichtenegger}, H. and {Magnes}, W. and {Nakamura}, R. and {Penz}, T. and {Schwingenschuh}, K. and {V{\"o}r{\"o}s}, Z. and {Zambelli}, W. and {Fornacon}, K. -H. and {Glassmeier}, K. -H. and {Richter}, I. and {Carr}, C. and {Kudela}, K. and {Shi}, J.~K. and {Zhao}, H. and {Motschmann}, U. and {Lebreton}, J. -P.},
        title = "{Magnetic field investigation of the Venus plasma environment: Expected new results from Venus Express}",
      journal = {\planss},
         year = 2006,
        month = nov,
       volume = {54},
       number = {13-14},
        pages = {1336-1343},
          doi = {10.1016/j.pss.2006.04.018},
       adsurl = {https://ui.adsabs.harvard.edu/abs/2006P&SS...54.1336Z}
}

@ARTICLE{Bothmer1998,
       author = {{Bothmer}, V. and {Schwenn}, R.},
        title = "{The structure and origin of magnetic clouds in the solar wind}",
      journal = {Annales Geophysicae},
         year = 1998,
        month = jan,
       volume = {16},
       number = {1},
        pages = {1-24},
          doi = {10.1007/s00585-997-0001-x},
       adsurl = {https://ui.adsabs.harvard.edu/abs/1998AnGeo..16....1B}
}

@ARTICLE{Kilpua2017,
       author = {{Kilpua}, Emilia and {Koskinen}, Hannu E.~J. and {Pulkkinen}, Tuija I.},
        title = "{Coronal mass ejections and their sheath regions in interplanetary space}",
      journal = {Living Reviews in Solar Physics},
         year = 2017,
        month = dec,
       volume = {14},
       number = {1},
          eid = {5},
        pages = {5},
          doi = {10.1007/s41116-017-0009-6},
       adsurl = {https://ui.adsabs.harvard.edu/abs/2017LRSP...14....5K}
}

@ARTICLE{Gosling1990,
       author = {{Gosling}, J.~T. and {Bame}, S.~J. and {McComas}, D.~J. and {Phillips}, J.~L.},
        title = "{Coronal mass ejections and large geomagnetic storms}",
      journal = {\grl},
         year = 1990,
        month = jun,
       volume = {17},
       number = {7},
        pages = {901-904},
          doi = {10.1029/GL017i007p00901},
       adsurl = {https://ui.adsabs.harvard.edu/abs/1990GeoRL..17..901G}
}

@ARTICLE{Al-Haddad2025,
       author = {{Al-Haddad}, Nada and {Lugaz}, No{\'e}},
        title = "{The Magnetic Field Structure of Coronal Mass Ejections: A More Realistic Representation}",
      journal = {\ssr},
         year = 2025,
        month = feb,
       volume = {221},
       number = {1},
          eid = {12},
        pages = {12},
          doi = {10.1007/s11214-025-01138-w},
       adsurl = {https://ui.adsabs.harvard.edu/abs/2025SSRv..221...12A}
}

@INPROCEEDINGS{Mostl2020,
       author = {{M{\"o}stl}, Christian and {Bailey}, Rachel L. and {Amerstorfer}, Ute V. and {Amerstorfer}, Tanja and {Weiss}, Andreas J. and {Reiss}, Martin A. and {Hinterreiter}, J{\"u}rgen and {Bauer}, Maike},
        title = "{Helio4Cast - a real time test environment to enhance space weather prediction at Earth}",
    booktitle = {EGU General Assembly Conference Abstracts},
         year = 2020,
       series = {EGU General Assembly Conference Abstracts},
        month = may,
          eid = {5202},
        pages = {5202},
          doi = {10.5194/egusphere-egu2020-5202},
       adsurl = {https://ui.adsabs.harvard.edu/abs/2020EGUGA..22.5202M}
}

@INPROCEEDINGS{Reiss2021,
       author = {{Reiss}, Martin and {Pevtsov}, Alexei and {Linker}, Jon and {Pinto}, Rui and {Arge}, Charles and {Muglach}, Karin and {Henney}, Carl J.},
        title = "{Activities related to the COSPAR ISWAT Cluster: Ambient Solar Magnetic Field, Heating and Spectral Irradiance}",
    booktitle = {43rd COSPAR Scientific Assembly. Held 28 January - 4 February},
         year = 2021,
       volume = {43},
        month = jan,
          eid = {p.2413},
        pages = {2413},
       adsurl = {https://ui.adsabs.harvard.edu/abs/2021cosp...43E2413R}
}

@INPROCEEDINGS{Weiss2021,
       author = {{Weiss}, Andreas J. and {Moestl}, Christian and {Nieves-Chinchilla}, Teresa and {Amerstorfer}, Ute and {Amerstorfer}, Tanja and {Reiss}, Martin and {Bailey}, Rachel and {Bauer}, Maike and {Palmerio}, Erika},
        title = "{Analysing coronal mass ejection flux ropes signatures using 3DCORE and Bayesian inference}",
    booktitle = {43rd COSPAR Scientific Assembly. Held 28 January - 4 February},
         year = 2021,
       volume = {43},
        month = jan,
          eid = {1050},
        pages = {1050},
       adsurl = {https://ui.adsabs.harvard.edu/abs/2021cosp...43E1050W}
}

@ARTICLE{Howard1982,
       author = {{Howard}, R.~A. and {Michels}, D.~J. and {Sheeley}, Jr., N.~R. and {Koomen}, M.~J.},
        title = "{The observation of a coronal transient directed at Earth.}",
      journal = {\apjl},
         year = 1982,
        month = dec,
       volume = {263},
        pages = {L101-L104},
          doi = {10.1086/183932},
       adsurl = {https://ui.adsabs.harvard.edu/abs/1982ApJ...263L.101H}
}

@ARTICLE{Jones2020,
       author = {{Jones}, S.~R. and {Scott}, C.~J. and {Barnard}, L.~A. and {Highfield}, R. and {Lintott}, C.~J. and {Baeten}, E.},
        title = "{The Visual Complexity of Coronal Mass Ejections Follows the Solar Cycle}",
      journal = {Space Weather},
         year = 2020,
        month = oct,
       volume = {18},
       number = {10},
          eid = {e02556},
        pages = {e02556},
          doi = {10.1029/2020SW00255610.1002/essoar.10503318.1},
       adsurl = {https://ui.adsabs.harvard.edu/abs/2020SpWea..1802556J}
}

@ARTICLE{Kaiser2008,
       author = {{Kaiser}, M.~L. and {Kucera}, T.~A. and {Davila}, J.~M. and {St. Cyr}, O.~C. and {Guhathakurta}, M. and {Christian}, E.},
        title = "{The STEREO Mission: An Introduction}",
      journal = {\ssr},
         year = 2008,
        month = apr,
       volume = {136},
       number = {1-4},
        pages = {5-16},
          doi = {10.1007/s11214-007-9277-0},
       adsurl = {https://ui.adsabs.harvard.edu/abs/2008SSRv..136....5K}
}

@ARTICLE{Chree1913,
       author = {{Chree}, C.},
        title = "{Some Phenomena of Sunspots and of Terrestrial Magnetism at Kew Observatory}",
      journal = {Philosophical Transactions of the Royal Society of London Series A},
         year = 1913,
        month = jan,
       volume = {212},
        pages = {75-116},
          doi = {10.1098/rsta.1913.0003},
       adsurl = {https://ui.adsabs.harvard.edu/abs/1913RSPTA.212...75C}
}

@ARTICLE{Sonnerup1967,
       author = {{Sonnerup}, B.~U.~O. and {Cahill}, Jr., L.~J.},
        title = "{Magnetopause Structure and Attitude from Explorer 12 Observations}",
      journal = {\jgr},
         year = 1967,
        month = jan,
       volume = {72},
        pages = {171},
          doi = {10.1029/JZ072i001p00171},
       adsurl = {https://ui.adsabs.harvard.edu/abs/1967JGR....72..171S}
}

@ARTICLE{Raouafi2023,
       author = {{Raouafi}, N.~E. and {Matteini}, L. and {Squire}, J. and {Badman}, S.~T. and {Velli}, M. and {Klein}, K.~G. and {Chen}, C.~H.~K. and {Matthaeus}, W.~H. and {Szabo}, A. and {Linton}, M. and {Allen}, R.~C. and {Szalay}, J.~R. and {Bruno}, R. and {Decker}, R.~B. and {Akhavan-Tafti}, M. and {Agapitov}, O.~V. and {Bale}, S.~D. and {Bandyopadhyay}, R. and {Battams}, K. and {Ber{\v{c}}i{\v{c}}}, L. and {Bourouaine}, S. and {Bowen}, T.~A. and {Cattell}, C. and {Chandran}, B.~D.~G. and {Chhiber}, R. and {Cohen}, C.~M.~S. and {D'Amicis}, R. and {Giacalone}, J. and {Hess}, P. and {Howard}, R.~A. and {Horbury}, T.~S. and {Jagarlamudi}, V.~K. and {Joyce}, C.~J. and {Kasper}, J.~C. and {Kinnison}, J. and {Laker}, R. and {Liewer}, P. and {Malaspina}, D.~M. and {Mann}, I. and {McComas}, D.~J. and {Niembro-Hernandez}, T. and {Nieves-Chinchilla}, T. and {Panasenco}, O. and {Pokorn{\'y}}, P. and {Pusack}, A. and {Pulupa}, M. and {Perez}, J.~C. and {Riley}, P. and {Rouillard}, A.~P. and {Shi}, C. and {Stenborg}, G. and {Tenerani}, A. and {Verniero}, J.~L. and {Viall}, N. and {Vourlidas}, A. and {Wood}, B.~E. and {Woodham}, L.~D. and {Woolley}, T.},
        title = "{Parker Solar Probe: Four Years of Discoveries at Solar Cycle Minimum}",
      journal = {\ssr},
         year = 2023,
        month = feb,
       volume = {219},
       number = {1},
          eid = {8},
        pages = {8},
          doi = {10.1007/s11214-023-00952-4},
archivePrefix = {arXiv},
       eprint = {2301.02727},
 primaryClass = {astro-ph.SR},
       adsurl = {https://ui.adsabs.harvard.edu/abs/2023SSRv..219....8R}
}

@ARTICLE{Jian2006,
       author = {{Jian}, L. and {Russell}, C.~T. and {Luhmann}, J.~G. and {Skoug}, R.~M.},
        title = "{Properties of Interplanetary Coronal Mass Ejections at One AU During 1995   2004}",
      journal = {\solphys},
         year = 2006,
        month = dec,
       volume = {239},
       number = {1-2},
        pages = {393-436},
          doi = {10.1007/s11207-006-0133-2},
       adsurl = {https://ui.adsabs.harvard.edu/abs/2006SoPh..239..393J}
}

@ARTICLE{Liu2005,
       author = {{Liu}, Y. and {Richardson}, J.~D. and {Belcher}, J.~W.},
        title = "{A statistical study of the properties of interplanetary coronal mass ejections from 0.3 to 5.4 AU}",
      journal = {\planss},
         year = 2005,
        month = jan,
       volume = {53},
       number = {1-3},
        pages = {3-17},
          doi = {10.1016/j.pss.2004.09.023},
       adsurl = {https://ui.adsabs.harvard.edu/abs/2005P&SS...53....3L}
}
\bibliographystyle{aasjournal}

\end{document}